\documentclass[sigconf,nonacm]{acmart}
\setcopyright{none}
\acmConference[]{}{}{}
\acmBooktitle{}
\acmDOI{}
\acmISBN{}

\usepackage{amsmath}
\usepackage{multirow}
\usepackage{stfloats}
\usepackage{pgfplots}
\usepgfplotslibrary{groupplots}
\pgfplotsset{compat=1.18}

\makeatletter
\renewcommand{\@titlefont}{\normalfont\bfseries\fontsize{16}{19}\selectfont}
\renewcommand{\@mkauthors@iii}{%
  \gdef\@currentauthors{}%
  \global\setbox\mktitle@bx=\vbox{%
    \unvbox\mktitle@bx
    \hsize=\textwidth\normalfont\normalsize\centering
    {\bfseries
      Qingrui Li\textsuperscript{1},\hspace{1pt}%
      Haowei Lou\textsuperscript{1},\hspace{1pt}%
      Chengkai Huang\textsuperscript{1,2,*},\hspace{1pt}%
      Quan Z. Sheng\textsuperscript{2},\hspace{1pt}%
      Lina Yao\textsuperscript{1}\par}
    \smallskip
    \textsuperscript{1}University of New South Wales, Sydney, NSW, Australia\par
    \textsuperscript{2}School of Computing, Macquarie University, Sydney, NSW, Australia\par
    \smallskip
    Emails: \href{mailto:qingrui.li@student.unsw.edu.au}{qingrui.li@student.unsw.edu.au},
    \href{mailto:haowei.lou@unsw.edu.au}{haowei.lou@unsw.edu.au},
    \href{mailto:chengkai.huang1@unsw.edu.au}{chengkai.huang1@unsw.edu.au}\par
    \href{mailto:michael.sheng@mq.edu.au}{michael.sheng@mq.edu.au},
    \href{mailto:lina.yao@unsw.edu.au}{lina.yao@unsw.edu.au}\par
    \medskip
  }%
}
\makeatother

\begin{document}

\title{When Do Pretrained Audio Representations Help Music Recommendation? A Study Across Ranking and Generative Interfaces}

\author{Qingrui Li}
\email{qingrui.li@student.unsw.edu.au}
\affiliation{\institution{University of New South Wales}\city{Sydney}\state{NSW}\country{Australia}}
\author{Haowei Lou}
\email{haowei.lou@unsw.edu.au}
\affiliation{\institution{University of New South Wales}\city{Sydney}\state{NSW}\country{Australia}}
\author{Chengkai Huang}
\authornote{Corresponding author.}
\email{chengkai.huang1@unsw.edu.au}
\affiliation{\institution{University of New South Wales}\city{Sydney}\state{NSW}\country{Australia}}
\affiliation{\department{School of Computing}\institution{Macquarie University}\city{Sydney}\state{NSW}\country{Australia}}
\author{Quan Z. Sheng}
\email{michael.sheng@mq.edu.au}
\affiliation{\department{School of Computing}\institution{Macquarie University}\city{Sydney}\state{NSW}\country{Australia}}
\author{Lina Yao}
\email{lina.yao@unsw.edu.au}
\affiliation{\institution{University of New South Wales}\city{Sydney}\state{NSW}\country{Australia}}
\renewcommand{\shortauthors}{Li et al.}

\begin{abstract}
Pretrained audio encoders are increasingly used to provide content representations for music recommendation, but their downstream value can change substantially with the recommendation interface through which they are used. We study this effect across six pretrained audio encoders and three representative settings: content-based retrieval, sequential recommendation, and Semantic-ID-based generative recommendation. Using matched random-feature and ID-only controls, we distinguish absolute recommendation performance from gains over content-free inputs. Across two music datasets and a full-catalog evaluation, music-pretrained and audio--text-aligned encoders are generally stronger when their pretrained geometry is used directly or converted into Semantic IDs, while these differences become much smaller after interaction-based adaptation in SASRec. We further examine how audio representations should be discretized for generative recommendation by varying residual-quantization width, depth, and retained layers. At epoch 10, depth one performs best at every tested width, while larger codebooks generally help at depth three; the best single-layer subset also outperforms the full three-layer identifier in every evaluated comparison. For CLAP-Music, the third-layer identifier ranks higher under both original and shuffled assignments, whereas the full three-layer identifier gains more from the original assignment. Better reconstruction likewise does not ensure better ranking. Overall, our results show that the utility of pretrained audio representations is interface-dependent and that Semantic-ID quality should be assessed by downstream ranking rather than representation fidelity alone.
\end{abstract}


\maketitle

\section{Introduction}

Audio embedding models provide the representational foundation for a broad range of music intelligence systems, including recommendation on online streaming services. By transforming a waveform into a compact vector, audio encoders make musical characteristics---such as timbre, rhythm, harmony, instrumentation, and higher-level semantics---accessible to downstream models.
The quality of this representation is therefore consequential not only for
music classification and understanding, but also for music recommendation: A 
recommender can make effective use of audio only when its item representations
retain information that helps distinguish and relate tracks in ways relevant
to listeners.

Audio representations are particularly valuable when behavioral evidence is
limited, since audio is available when a track enters the catalog. They provide
side information for music recommendation and item cold-start
\cite{oord2013deep,chen2021audio}. Their usefulness depends on what the audio
encoder preserves.

Recent large-scale pretraining has produced increasingly capable audio
representation models. Speech-oriented models such as Wav2Vec~2.0 and HuBERT,
music-oriented models such as Music2Vec and MERT, and language--audio models
such as CLAP learn from different data domains and objectives
\cite{baevski2020wav2vec,hsu2021hubert,li2022music2vec,li2024mert,elizalde2023clap}.
These models are commonly evaluated on audio classification, tagging, music
understanding, or cross-modal retrieval tasks. Such evaluations establish that
an embedding contains useful acoustic or semantic information, but they do not
establish that its geometry is suitable for recommendation. Recommendation is
relational and behavior-dependent: two tracks that share a class label need not
be interchangeable for a listener, while tracks that differ in conventional
music-understanding labels may still occur in the same preference context.
Consequently, performance on standard audio benchmarks may not transfer predictably to recommendations.

A small body of work has begun to evaluate pretrained audio embeddings in
music recommender systems. For example, \citet{tamm2026adopting} compare pretrained audio representations across five recommendation approaches, including KNN and BERT4Rec~\cite{sun2019bert4rec}, in hot- and cold-start settings, with shuffled-embedding controls. This provides important evidence for conventional and sequential
recommendation, but does not cover the emerging generative recommendation
(GenRec) paradigm. GenRec changes the role of the content representation: rather than using an audio vector only for similarity or as a model input,
methods such as TIGER quantize a continuous item representation into a sequence
of discrete Semantic-ID tokens and autoregressively generate the next item
\cite{rajput2023tiger}. Recent generative music recommendation has also explored
multimodal Semantic IDs that include audio features \cite{kim2026fusid}, but it
does not provide a controlled comparison of alternative pretrained audio
encoders across conventional and generative recommenders. It therefore remains unclear how the choice of pretrained audio encoder
and its quantization affect generative recommendation.

We extend pretrained audio representation comparison to TIGER and examine how quantization depth, retained RVQ layers, and the original item--SID assignment affect recommendation. We benchmark six encoders across content-based nearest-neighbor retrieval with KNN, self-attentive sequential recommendation with SASRec, and Semantic-ID-based generative recommendation with TIGER. The selected encoders span speech, music, and general-audio pretraining as well as masked prediction, teacher--student, and audio--text contrastive objectives. We vary residual-quantization codebook width and depth and compare Semantic IDs formed from different retained RVQ-layer subsets. We complement the encoder comparison with dimension-matched random features and an ID-only SASRec baseline. For CLAP-Music, we also compare P3 and P123 under original and shuffled item--SID assignments. These controls separate overall ranking performance, gains over content-free inputs, and sensitivity to the assignment of items to structured IDs.

The main contributions of this work are summarized as follows:
\begin{itemize}
    \item Across six encoders and three recommendation interfaces, we distinguish absolute ranking from gains over matched content-free controls: SASRec ranks strongest, yet its audio gains are modest on the two smaller catalogs.
    \item Under the epoch-10 protocol, depth one ranks best at every tested width, while larger codebooks generally help in the depth-three sweep. Selected single-layer subsets outperform the full three-layer identifier in all evaluated comparisons.
    \item For CLAP-Music, stronger ranking need not coincide with better audio reconstruction or a larger original-assignment gain: P3 ranks above P123 under both assignments, while P123 benefits more from the original item--SID mapping.
\end{itemize}

\section{Related Work}
\subsection{Pretrained Audio Representation Models}

Self-supervised audio encoders differ substantially in both their pretraining data and training objective. Wav2Vec~2.0 masks latent speech features and uses a
contrastive objective to identify quantized targets among distractors
\cite{baevski2020wav2vec}. HuBERT instead constructs discrete targets through
offline clustering and predicts the hidden units at masked time steps
\cite{hsu2021hubert}. Both models were developed primarily for speech, making
them useful controls for testing how speech-derived acoustic representations
transfer to music.

Music2Vec adapts the data2vec teacher--student framework~\cite{baevski2022data2vec} to raw music: a student
receives a masked input and regresses contextualized targets produced by an
exponential-moving-average teacher \cite{li2022music2vec}. MERT introduces
music-specific masked prediction with complementary teachers: an acoustic
teacher based on residual vector quantization and a musical teacher derived
from constant-Q-transform features \cite{li2024mert}. These objectives are
intended to encode aspects of musical audio that are not central to
speech-oriented pretraining.
CLMR learns music representations by contrasting augmented audio views and evaluates them on music classification~\cite{spijkervet2021clmr}.

Contrastive Language--Audio Pretraining (CLAP) aligns an audio encoder and a
text encoder in a shared space using paired audio and natural-language
descriptions \cite{elizalde2023clap}. Large-scale CLAP variants expand the
training mixture and introduce mechanisms such as global--local feature fusion
and keyword-to-caption augmentation \cite{wu2023laionclap}. Compared with
masked-prediction objectives, CLAP directly encourages high-level semantic
alignment, but that alignment need not coincide with behavioral similarity.
We include both general-audio and music-oriented CLAP checkpoints to examine
this distinction. MuLan also learns a joint space from music audio and free-form text, evaluated through tagging and cross-modal retrieval~\cite{huang2022mulan}. 

\subsection{Audio-Aware Music Recommendation}

Content-based recommendation uses item attributes to complement or replace
interaction-derived representations. In music, \citet{oord2013deep} trained a
convolutional network to predict collaborative latent factors from audio,
thereby enabling recommendations for tracks without sufficient usage data.
\citet{chen2021audio} learned an audio embedding through metric learning,
anchoring liked and disliked tracks to user representations derived from
listening behavior. These studies demonstrate that the recommendation utility
of audio can be learned explicitly from interaction signals.

Our setting is complementary: rather than proposing a new task-specific audio
encoder, we ask how reusable pretrained encoders behave when inserted into
modern recommendation pipelines. A cosine nearest-neighbor method provides a
transparent measurement of the original embedding geometry. SASRec provides a
strong sequential baseline that applies causal self-attention to an interaction
history and predicts the next item \cite{kang2018sasrec}. Evaluating both
allows us to distinguish immediately available content similarity from the
representation after adaptation to behavior.

\subsection{Generative Recommendation and Semantic IDs}

\begingroup\emergencystretch=1em\relax
Conventional recommenders score candidates represented by atomic item IDs or
dense vectors. Generative retrieval instead represents a target as a sequence
of discrete tokens and directly decodes that sequence~\cite{tay2022dsi}. TIGER
\cite{rajput2023tiger} uses a residual-quantized autoencoder to map a continuous
content embedding to a tuple of codewords, called a Semantic ID, and trains an
encoder--decoder Transformer to generate the next item's tuple from a user's
history.
\par\endgroup

Residual quantization is central to this representation. A standard vector
quantizer selects one codeword from a codebook \cite{oord2017vqvae}. Residual
quantization applies multiple codebooks successively: the first codeword
approximates the input, and each later codeword approximates the remaining
residual \cite{lee2022rqvae}. We refer to the
number of codewords per codebook as \emph{width} and the number of successive
codebooks as \emph{depth}. Width controls the branching factor at each token
position, whereas depth controls identifier length and progressive
refinement. These choices jointly affect reconstruction error, code
utilization, collisions, sequence length, and decoding complexity. Prior work
establishes Semantic IDs as an effective interface between content and
generative retrieval \cite{rajput2023tiger}. \citet{mei2025semanticids} compare content-based and random Semantic IDs for music recommendation and study the trade-off between codebook size and model capacity. Our study focuses on how this interface behaves
when constructed from different pretrained audio spaces.

\section{Method}
\label{sec:method}

\begin{figure*}[!t]
\centering
\includegraphics[width=\textwidth]{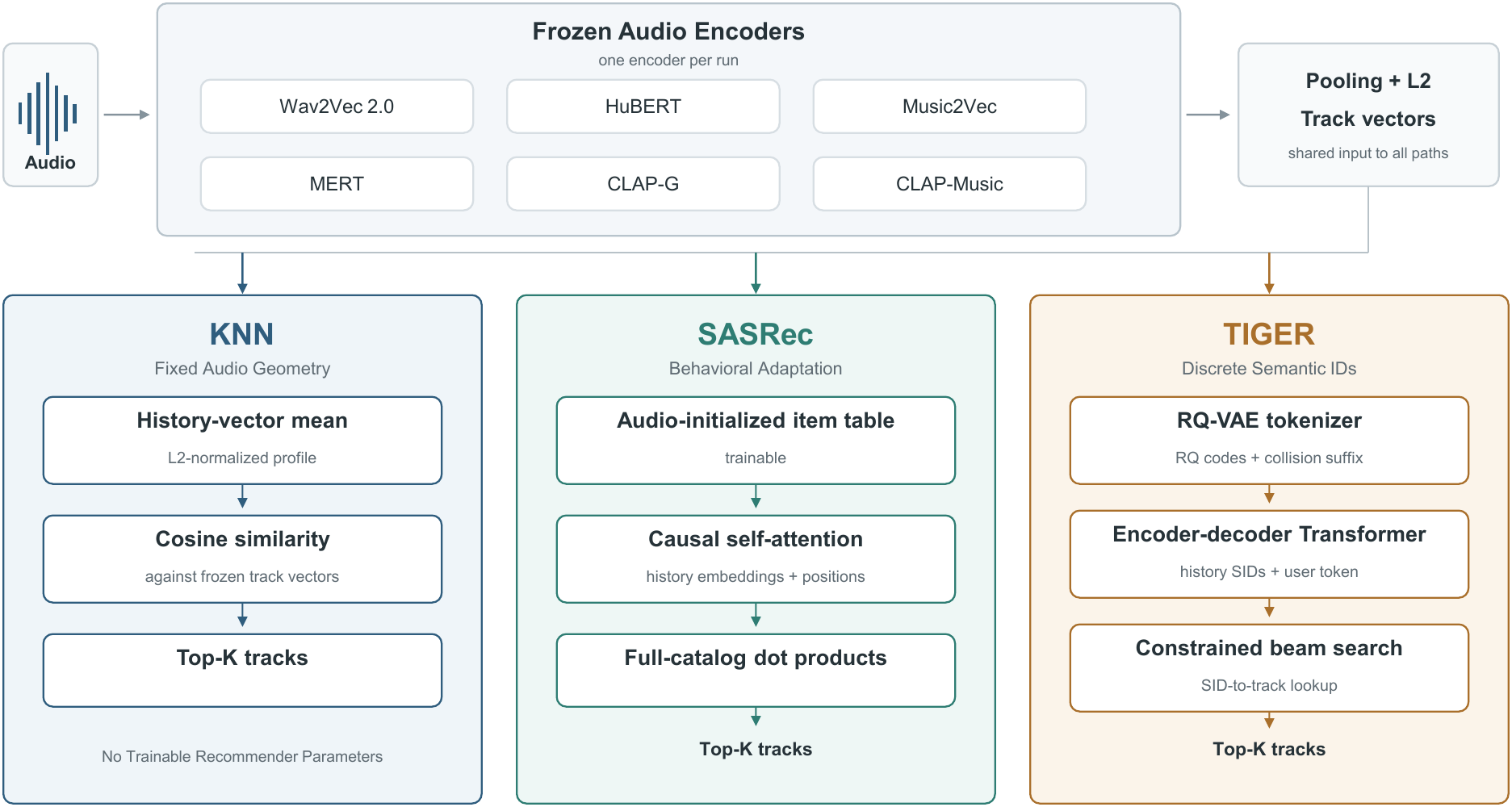}
\caption{Overview of the three recommendation interfaces. Each frozen audio encoder produces pooled, normalized track vectors. KNN uses the original audio geometry; SASRec adapts an audio-initialized item table through sequence modeling; TIGER quantizes the vectors into Semantic IDs for constrained generation. The tokenizer is trained offline; Semantic IDs remain fixed during generator training.}
\label{fig:recommendation-interfaces}
\Description{Six individually boxed alternative frozen audio encoders feed three parallel paths: history-averaged cosine retrieval in KNN, a trainable audio-initialized item table and causal attention in SASRec, and residual quantization followed by history-conditioned generation and constrained beam search in TIGER. All paths return ranked tracks.}
\end{figure*}

\subsection{Problem Formulation}

Let $\mathcal{I}$ denote the track catalog and
$S_u=(i_{u,1},\ldots,i_{u,T_u})$ the chronologically ordered
interaction sequence of user $u$. Given an observed prefix
$S_{u,\leq t}$, the task is to rank the held-out next track
$i_{u,t+1}$ among the catalog. Each track $i$ has an audio waveform
$x_i$. A pretrained audio encoder $f_\theta$ produces an offline
content representation
\begin{equation}
    \mathbf{z}_i =
    \operatorname{norm}\!\left(
        \operatorname{pool}\!\left(f_\theta(x_i)\right)
    \right)
    \in \mathbb{R}^{d_\theta},
    \label{eq:audio-embedding}
\end{equation}
where $\operatorname{pool}(\cdot)$ is the encoder-specific temporal
and layer aggregation, and $\operatorname{norm}(\cdot)$ denotes
row-wise $\ell_2$ normalization. The audio encoder and extraction
pipeline remain frozen. For a given encoder, the same precomputed
track vectors are supplied to KNN, SASRec, and TIGER. Trainable audio
representations below therefore refer only to SASRec's downstream
item-embedding table, not to the pretrained encoder.

\subsection{Recommendation Interfaces}

Figure~\ref{fig:recommendation-interfaces} summarizes how the same pretrained
track representations are used for direct retrieval, sequential adaptation,
and generative retrieval.

\paragraph{Content-based nearest neighbors}
Following prior work that evaluates pretrained audio representations with KNN \cite{tamm2026adopting}, we use KNN to probe the original geometry of the pretrained audio space
without behavioral adaptation. Let $\mathcal{H}_{u,t}$ be the tracks
observed before the prediction target. We form the user vector and
candidate score as
\begin{equation}
    \mathbf{q}_{u,t} =
    \operatorname{norm}\!\left(
        \frac{1}{|\mathcal{H}_{u,t}|}
        \sum_{j\in\mathcal{H}_{u,t}}\mathbf{z}_j
    \right),
    \qquad
    s_{\mathrm{KNN}}(u,i)=\mathbf{q}_{u,t}^{\top}\mathbf{z}_i .
    \label{eq:knn}
\end{equation}
Because both vectors are normalized, this score is cosine
similarity. Tracks in the most recent 100 interactions are removed before ranking.

\begingroup\emergencystretch=1em\relax
\paragraph{Sequential adaptation with SASRec}
SASRec applies causal self-attention to the ordered interaction
sequence \cite{kang2018sasrec}. Its item-embedding table is initialized
directly from the audio vectors,
$\mathbf{e}_i^{(0)}=\mathbf{z}_i$, and remains trainable together with
the sequence model. The hidden dimension is matched to
$d_\theta$, avoiding an additional projection bottleneck. Given
positional embeddings $\mathbf{p}_\tau$, SASRec produces
\begin{equation}
    (\mathbf{h}_{u,1},\ldots,\mathbf{h}_{u,t})
    =
    F_{\mathrm{SASRec}}\!\left(
        \mathbf{e}_{i_{u,1}}+\mathbf{p}_1,\ldots,
        \mathbf{e}_{i_{u,t}}+\mathbf{p}_t
    \right).
    \label{eq:sasrec}
\end{equation}
The final state scores a candidate through
$\mathbf{h}_{u,t}^{\top}\mathbf{e}_i$, and the model is trained with
full-softmax next-item cross-entropy. SASRec can therefore reshape the
initial audio geometry using interaction supervision.
\par\endgroup

\paragraph{Generative retrieval with TIGER}
TIGER converts each continuous audio vector into a discrete Semantic
ID before modeling the user sequence \cite{rajput2023tiger}. An
encoder adapter first maps $\mathbf{z}_i$ to a latent
$\mathbf{h}_i$. Let $\mathbf{r}_i^{(0)}=\mathbf{h}_i$ and
$\mathcal{C}^{(\ell)}
=\{\mathbf{c}^{(\ell)}_0,\ldots,\mathbf{c}^{(\ell)}_{W-1}\}$
be the codebook at residual level $\ell$. Hard quantization selects
\begin{align}
    k_i^{(\ell)}
    &=
    \arg\min_{k\in\{0,\ldots,W-1\}}
    \left\|
        \mathbf{r}_i^{(\ell-1)}-\mathbf{c}^{(\ell)}_k
    \right\|_2^2,
    \label{eq:rq-assignment}\\
    \mathbf{r}_i^{(\ell)}
    &=
    \mathbf{r}_i^{(\ell-1)}
    -\mathbf{c}^{(\ell)}_{k_i^{(\ell)}} .
    \label{eq:rq-residual}
\end{align}
The quantized latent is
$\widehat{\mathbf{h}}_i
=\sum_{\ell=1}^{D}\mathbf{c}^{(\ell)}_{k_i^{(\ell)}}$.
The tokenizer is trained using reconstruction, codebook, and
commitment losses.

Tracks may share all $D$ quantization codes. We therefore assign a
deterministic collision suffix $c_i$ within each shared-code group:
\begin{equation}
    \operatorname{SID}(i)
    =
    \bigl(k_i^{(1)},\ldots,k_i^{(D)},c_i\bigr),
    \label{eq:sid}
\end{equation}
yielding a catalog-unique identifier. Let
$\mathbf{y}_{u,t+1}=\operatorname{SID}(i_{u,t+1})$. The
encoder--decoder Transformer minimizes
\begin{equation}
    \mathcal{L}_{\mathrm{TIGER}}
    =
    -\sum_{m=1}^{D+1}
    \log p\!\left(
        y_{u,t+1,m}
        \mid
        y_{u,t+1,<m},
        \tau_u,
        \operatorname{SID}(\mathcal{H}_{u,t})
    \right),
    \label{eq:tiger-objective}
\end{equation}
where $\tau_u$ is the user token. At inference, constrained beam
search expands only prefixes that can terminate in a valid catalog
identifier. Completed identifiers are mapped back to tracks,
tracks in the last 100 interactions are removed, and candidates are ranked by
generation score.

\subsection{Semantic-ID Interventions}

We study codebook width and residual depth separately. The width
intervention uses
$W\in\{64,256,1024,4096\}$ with $D=3$, whereas the depth intervention
uses $D\in\{1,3,6,12\}$ at the same four widths. Each width or depth condition
trains its corresponding tokenizer and generator independently while
holding the remaining protocol fixed. These sweeps use suffix capacity
$P=4096$, TIGER learning rate $3\times10^{-4}$, and seed 2020;
results are reported on the validation split after 10 epochs.

The retained-layer intervention reuses a fixed reference tokenizer
with $W=256$ and $D=3$ for each dataset--encoder pair. We evaluate all seven nonempty layer subsets,
denoted P1, P2, P3, P12, P13, P23, and P123; the digits identify the
retained RQ-code positions. The three cumulative-prefix cases are
\begin{align}
    \mathrm{P1}(i)
    &=
    [k_i^{(1)},\mathrm{NULL},\mathrm{NULL},c_i^{(1)}],\\
    \mathrm{P12}(i)
    &=
    [k_i^{(1)},k_i^{(2)},\mathrm{NULL},c_i^{(12)}],\\
    \mathrm{P123}(i)
    &=
    [k_i^{(1)},k_i^{(2)},k_i^{(3)},c_i^{(123)}].
\end{align}
For each proper subset, the suffix is reassigned deterministically in
canonical item order within each retained-code bucket; P123 reuses
the reference tokenizer's original suffix exactly. Inactive positions contain a fixed
NULL token that is excluded from the training loss and inserted
during constrained decoding without branching or score accumulation.
All conditions retain four slots: three RQ-code positions and one
collision-disambiguation suffix with capacity $P=4096$.
A separate generator is freshly trained for every feasible subset from
the same initialization for 10 epochs, with learning rate $3\times10^{-4}$
and seed 2020. We report epoch-10 full-validation scores; configurations
that exceed the fixed suffix capacity are marked unavailable.
The comparison evaluates the complete Semantic IDs formed from each
retained-layer subset and its suffix, with a separately trained generator.

\begingroup\emergencystretch=2em\relax
\subsection{Content and Item--SID Controls}
\label{sec:controls}
\paragraph{Random features and ID-only recommendation}
We construct standard Gaussian random feature banks with dimensions 512 and 768, normalize each track vector to unit length, and share each bank across KNN, SASRec, and TIGER. KNN uses these vectors directly; SASRec uses them to initialize its trainable item table; TIGER learns a new $W=256,D=3$ tokenizer from each bank before training its generator. Dimension-matched audio and random inputs use the same recommender architecture and training budget. The ID-only control uses the same SASRec architecture with its default trainable item-ID initialization and no external feature bank. This comparison distinguishes gains associated with audio input from performance supported by item identities and interaction sequences.

\paragraph{Original and shuffled Semantic IDs}
For CLAP-Music on each smaller catalog, we reuse the fixed reference tokenizer from the retained-layer experiment and cross P3/P123 with original/shuffled item--SID assignments. Shuffling permutes complete SID rows, including the suffix, across items. It preserves the catalog's SID multiset, codeword frequencies, collision-bucket sizes, and valid code paths, while changing their association with item identities and interaction sequences. Within each dataset and seed, both subsets use the same item permutation, and all four generators share initial weights and training-example order. P3 retains the third RQ code and suffix, with two fixed NULL positions; P123 uses all four tokens. Training retains the original retained-layer budget and averages token loss over active positions.

For a metric $M$, we measure assignment gains as $G_S=M(S,\mathrm{original})-M(S,\mathrm{shuffled})$, for $S\in\{\mathrm{P3},\mathrm{P123}\}$. The interaction $I=G_{\mathrm{P3}}-G_{\mathrm{P123}}$ compares these gains within complete four-condition seed blocks. This separates a subset's absolute ranking advantage from its benefit from the original item--SID association.
\par\endgroup

\section{Experiment Settings}

\begin{table}[!t]
\centering
\small
\setlength{\tabcolsep}{3.5pt}
\caption{Canonical audio-embedding settings used on both datasets.
All final track vectors are row-wise $\ell_2$-normalized.}
\label{tab:audio-encoders}

\begin{tabular}{l c c p{3.6cm}}
\toprule
Embedder & Frequency & Dim. & Pooling \\
\midrule

Wav2Vec~2.0
& 16 kHz  & 768
& Temporal mean of the final hidden state \\

HuBERT
& 16 kHz & 768
& Temporal mean of the final hidden state \\

Music2Vec
& 16 kHz & 768
& Mean of each of 13 hidden states, followed by a raw-layer mean \\

MERT
& 24 kHz  & 768
& Mean of each of 13 hidden states, followed by a raw-layer mean \\

CLAP-G
& 48 kHz  & 512
& CLAP-Audio with fusion enabled \\

CLAP-Music
& 48 kHz  & 512
& CLAP-Audio with fusion disabled and HTSAT-base \\

\bottomrule
\end{tabular}
\end{table}

\begin{table*}[!t]
\caption{Test performance at $K=50$ across audio embeddings, random-feature controls, and recommender systems. Results are mean $\pm$ sample standard deviation across runs; KNN variation for audio inputs measures execution repeatability. Random controls use seeds 2020--2024; suffixes 512 and 768 denote input dimensions, matching CLAP and the other encoders, respectively. ID-only SASRec results are in Table~\ref{tab:id-only}. \textbf{Bold} and \underline{underlined} values denote the best and second-best audio embeddings within each dataset, recommender, and metric.}
\label{tab:main-results}
    \centering
    \scriptsize
    \setlength{\tabcolsep}{1.2pt}
    \renewcommand{\arraystretch}{1.10}
    \begin{tabular*}{\textwidth}{@{\extracolsep{\fill}}lrrrrrrrrr@{}}
    \toprule
    & \multicolumn{3}{c}{\textbf{KNN}}
    & \multicolumn{3}{c}{\textbf{SASRec}}
    & \multicolumn{3}{c}{\textbf{TIGER}} \\
    \cmidrule(lr){2-4}\cmidrule(lr){5-7}\cmidrule(lr){8-10}
    \textbf{Input}
    & \multicolumn{1}{c}{\textbf{R@50}$\uparrow$}
    & \multicolumn{1}{c}{\textbf{N@50}$\uparrow$}
    & \multicolumn{1}{c}{\textbf{MRR@50}$\uparrow$}
    & \multicolumn{1}{c}{\textbf{R@50}$\uparrow$}
    & \multicolumn{1}{c}{\textbf{N@50}$\uparrow$}
    & \multicolumn{1}{c}{\textbf{MRR@50}$\uparrow$}
    & \multicolumn{1}{c}{\textbf{R@50}$\uparrow$}
    & \multicolumn{1}{c}{\textbf{N@50}$\uparrow$}
    & \multicolumn{1}{c}{\textbf{MRR@50}$\uparrow$} \\
    \midrule
    \multicolumn{10}{@{}l}{\textbf{LFM-AM-4996}} \\
Random-512 & $0.0075\pm0.0007$ & $0.0018\pm0.0002$ & $0.0005\pm0.0001$ & $0.4618\pm0.0028$ & $0.2879\pm0.0004$ & $0.2429\pm0.0007$ & $0.2777\pm0.0048$ & $0.1860\pm0.0038$ & $0.1626\pm0.0042$ \\
Random-768 & $0.0079\pm0.0008$ & $0.0020\pm0.0002$ & $0.0007\pm0.0001$ & $0.4569\pm0.0042$ & $0.2856\pm0.0014$ & $0.2413\pm0.0013$ & $0.2933\pm0.0114$ & $0.1938\pm0.0066$ & $0.1686\pm0.0056$ \\
\addlinespace[2pt]
    Wav2Vec2 & 0.0190$\pm$0.0000 & 0.0047$\pm$0.0000 & 0.0015$\pm$0.0000 & 0.4661$\pm$0.0015 & 0.2875$\pm$0.0003 & 0.2414$\pm$0.0007 & 0.3067$\pm$0.0084 & 0.1979$\pm$0.0039 & 0.1704$\pm$0.0029 \\
    Music2Vec & \underline{0.0280$\pm$0.0000} & \underline{0.0077$\pm$0.0000} & \underline{0.0030$\pm$0.0000} & 0.4675$\pm$0.0010 & 0.2875$\pm$0.0011 & 0.2411$\pm$0.0013 & 0.3142$\pm$0.0052 & 0.2023$\pm$0.0034 & 0.1740$\pm$0.0031 \\
    HuBERT & 0.0193$\pm$0.0000 & 0.0052$\pm$0.0000 & 0.0020$\pm$0.0000 & 0.4632$\pm$0.0006 & 0.2871$\pm$0.0009 & 0.2418$\pm$0.0012 & 0.3089$\pm$0.0125 & 0.1989$\pm$0.0066 & 0.1709$\pm$0.0051 \\
    MERT & 0.0195$\pm$0.0000 & 0.0052$\pm$0.0000 & 0.0019$\pm$0.0000 & 0.4630$\pm$0.0017 & 0.2869$\pm$0.0007 & 0.2414$\pm$0.0006 & 0.3067$\pm$0.0070 & 0.1963$\pm$0.0031 & 0.1683$\pm$0.0024 \\
    CLAP-G & 0.0265$\pm$0.0000 & 0.0073$\pm$0.0000 & 0.0029$\pm$0.0000 & \underline{0.4696$\pm$0.0035} & \underline{0.2905$\pm$0.0011} & \textbf{0.2444$\pm$0.0009} & \underline{0.3159$\pm$0.0068} & \underline{0.2032$\pm$0.0024} & \underline{0.1746$\pm$0.0013} \\
    CLAP-Music & \textbf{0.0369$\pm$0.0000} & \textbf{0.0098$\pm$0.0000} & \textbf{0.0036$\pm$0.0000} & \textbf{0.4701$\pm$0.0027} & \textbf{0.2906$\pm$0.0005} & \underline{0.2443$\pm$0.0008} & \textbf{0.3282$\pm$0.0120} & \textbf{0.2080$\pm$0.0065} & \textbf{0.1774$\pm$0.0050} \\

    \midrule
    \multicolumn{10}{@{}l}{\textbf{Onion-Dedup-5000}} \\
Random-512 & $0.0087\pm0.0008$ & $0.0021\pm0.0002$ & $0.0006\pm0.0001$ & $0.4589\pm0.0012$ & $0.2755\pm0.0006$ & $0.2286\pm0.0007$ & $0.2747\pm0.0092$ & $0.1775\pm0.0067$ & $0.1533\pm0.0062$ \\
Random-768 & $0.0068\pm0.0004$ & $0.0016\pm0.0001$ & $0.0004\pm0.0000$ & $0.4531\pm0.0027$ & $0.2734\pm0.0010$ & $0.2274\pm0.0014$ & $0.2846\pm0.0160$ & $0.1844\pm0.0120$ & $0.1594\pm0.0110$ \\
\addlinespace[2pt]
    Wav2Vec2 & 0.0169$\pm$0.0000 & 0.0046$\pm$0.0000 & 0.0018$\pm$0.0000 & 0.4624$\pm$0.0027 & 0.2761$\pm$0.0015 & 0.2285$\pm$0.0016 & 0.2944$\pm$0.0050 & 0.1840$\pm$0.0011 & 0.1565$\pm$0.0020 \\
    Music2Vec & 0.0235$\pm$0.0000 & 0.0061$\pm$0.0000 & 0.0021$\pm$0.0000 & 0.4604$\pm$0.0021 & 0.2750$\pm$0.0011 & 0.2277$\pm$0.0015 & 0.3069$\pm$0.0085 & 0.1898$\pm$0.0043 & 0.1606$\pm$0.0040 \\
    HuBERT & 0.0194$\pm$0.0000 & 0.0052$\pm$0.0000 & 0.0019$\pm$0.0000 & 0.4601$\pm$0.0025 & 0.2738$\pm$0.0008 & 0.2262$\pm$0.0010 & 0.2961$\pm$0.0103 & 0.1861$\pm$0.0048 & 0.1586$\pm$0.0036 \\
    MERT & 0.0174$\pm$0.0000 & 0.0048$\pm$0.0000 & 0.0019$\pm$0.0000 & 0.4588$\pm$0.0026 & 0.2744$\pm$0.0020 & 0.2273$\pm$0.0024 & 0.3020$\pm$0.0084 & 0.1874$\pm$0.0041 & 0.1589$\pm$0.0033 \\
    CLAP-G & \underline{0.0271$\pm$0.0000} & \underline{0.0069$\pm$0.0000} & \underline{0.0023$\pm$0.0000} & \underline{0.4661$\pm$0.0034} & \underline{0.2778$\pm$0.0010} & \underline{0.2297$\pm$0.0007} & \underline{0.3216$\pm$0.0017} & \underline{0.1969$\pm$0.0016} & \underline{0.1658$\pm$0.0017} \\
    CLAP-Music & \textbf{0.0439$\pm$0.0000} & \textbf{0.0119$\pm$0.0000} & \textbf{0.0046$\pm$0.0000} & \textbf{0.4677$\pm$0.0024} & \textbf{0.2784$\pm$0.0010} & \textbf{0.2301$\pm$0.0013} & \textbf{0.3216$\pm$0.0054} & \textbf{0.1978$\pm$0.0030} & \textbf{0.1669$\pm$0.0025} \\

    \bottomrule
    \end{tabular*}

\end{table*}
\label{sec:experiment-settings}

\subsection{Datasets}
We use two music recommendation datasets: \textbf{LFM1b}~\cite{schedl2016lfm} and
\textbf{Music4All-Onion}~\cite{moscati2022onion}.
Music4All-Onion extends Music4All~\cite{santana2020music4all} with user
listening records and additional content features. Due to the high cost of retrieving
and matching audio for large-scale catalogs, we construct subsets of approximately
5,000 tracks from each dataset. We sample 10,000 users with at least 21
interaction records (seed 42), remove tracks with fewer than five records
among these users, and retain the 5,000 most frequent tracks. We then retain
users with at least 20 remaining records. Removing four tracks absent from
the LFM item metadata yields LFM-AM-4996. For Onion, these counts use unique
user--track pairs; each retained pair uses its last listening timestamp,
with repeat counts stored separately. The resulting LFM subset contains 7,987 users, 4,996 tracks, and
$1{,}904{,}242$ unique user--track interactions, while the Onion subset
contains 8,855 users, 5,000 tracks, and $1{,}972{,}371$ unique interactions.
Audio is available for all retained tracks. An item-to-audio mapping links
each track to its waveform and fixes the row order of its audio embeddings.
These subsets support the encoder comparisons and Semantic-ID studies; the full-catalog evaluation extends the encoder and content-control comparisons (Appendix~\ref{app:full-catalog-results}).

\begingroup\emergencystretch=1em\relax
For both datasets, interactions are chronologically ordered and evaluated
using a leave-one-out protocol: the final interaction is used for testing, the
penultimate interaction for validation, and the remaining interactions for
training. For the main, full-catalog, and content-control comparisons, all three methods share the users, targets, splits, and candidate catalog. Test histories include the validation interaction, and tracks in the last 100 interactions are excluded from the candidates. We report Recall@50, NDCG@50~\cite{jarvelin2002gain}, and MRR@50.
\par\endgroup

\paragraph{Audio encoders}
Once the waveform file for each track is retrieved, we apply a standardized
pipeline to extract embeddings from each audio encoder. For Wav2Vec~2.0~\cite{baevski2020wav2vec},
HuBERT~\cite{hsu2021hubert}, and Music2Vec~\cite{li2022music2vec}, we use
30-second, 16-kHz mono audio, while MERT~\cite{li2024mert} uses 5-second,
24-kHz mono audio. Longer clips are center-cropped, whereas shorter clips are
zero-padded. Wav2Vec~2.0 and HuBERT use mean pooling over the final hidden
states, while Music2Vec and MERT aggregate representations across both time
and hidden layers. For CLAP~\cite{wu2023laionclap}, we use 48-kHz mono audio
and the official 10-second preprocessing. For longer recordings, the
general-audio model (\textbf{CLAP-G}) uses feature fusion, while the
music-oriented model (\textbf{CLAP-Music}) uses a random 10-second crop.
Shorter recordings are repeated and zero-padded. All embeddings are
$\ell_2$-normalized before downstream evaluation. Full details of the embedding
extraction settings are provided in Table~\ref{tab:audio-encoders}.
Additionally, we employ residual vector quantization (\textbf{RVQ})~\cite{zeghidour2021soundstream}
to discretize continuous audio embeddings into hierarchical Semantic IDs.

\paragraph{Model configurations}
For each recommender system, we follow a standardized training and evaluation pipeline. We adapt the trainable item representations of SASRec~\cite{kang2018sasrec} and the content-derived Semantic IDs of TIGER~\cite{rajput2023tiger} to the same pretrained audio vectors. Each method retains its own model configuration, while the main comparisons share the data splits, targets, and candidate filtering described above.

KNN performs exact top-50 cosine retrieval using a mean profile over the full available history, without trainable parameters. SASRec initializes a trainable item table from the audio embeddings and uses Adam~\cite{kingma2015adam} with learning rate $10^{-4}$. TIGER uses a three-layer reference configuration with $W=256$ and suffix capacity $P=4096$. Its generator uses AdamW~\cite{loshchilov2019adamw} with learning rate $3\times10^{-4}$ and weight decay $0.01$. Both learned recommenders use up to 100 recent tracks and are evaluated at epoch 10.

\begingroup\emergencystretch=1em\relax
Within each recommender, the training protocol is held fixed across encoders. The main, full-catalog, and control comparisons report test performance; the width/depth and seven-subset interventions report validation performance. Random-feature and ID-only controls use seeds 2020--2024 on the two smaller catalogs and seed 2020 on Full. Control tables report means and sample standard deviations, with run counts stated for each comparison. Differences are paired by dataset, dimension, method, and seed; interaction estimates use the intersection of all four conditions. Appendix~\ref{app:configuration-details} consolidates the architecture, optimization, tokenizer selection, and experiment-specific settings. Computational resources included the Katana cluster~\cite{unsw2010katana}.
\par\endgroup

\begin{table}[!t]
\centering
\caption{NDCG@50 for retained RVQ-layer subsets at epoch 10 on validation. Digits identify the retained layers. Bold and underlined values denote the best and second-best subsets within each encoder and dataset. Recall@50 and MRR@50 are in Appendix~\ref{app:semantic-id-results}.}
\label{tab:prefix}
\footnotesize
\setlength{\tabcolsep}{1.2pt}
\renewcommand{\arraystretch}{1.03}
\begin{tabular*}{\columnwidth}{@{\extracolsep{\fill}}lrrrrrrr@{}}
\toprule
Embedding & P1 & P2 & P3 & P12 & P13 & P23 & P123 \\
\midrule
\multicolumn{8}{l}{\textbf{LFM-AM-4996}} \\
Wav2Vec2 & \underline{0.2697} & 0.2540 & \textbf{0.2793} & 0.2510 & 0.2487 & 0.2431 & 0.2420 \\
Music2Vec & \textbf{0.2687} & 0.2652 & \underline{0.2665} & 0.2473 & 0.2387 & 0.2443 & 0.2394 \\
HuBERT & \underline{0.2657} & 0.2529 & \textbf{0.2801} & 0.2449 & 0.2625 & 0.2451 & 0.2448 \\
MERT & 0.2609 & \textbf{0.2746} & \underline{0.2656} & 0.2488 & 0.2433 & 0.2565 & 0.2365 \\
CLAP-G & \underline{0.2548} & 0.2484 & \textbf{0.2656} & 0.2304 & 0.2375 & 0.2342 & 0.2323 \\
CLAP-Music & 0.2565 & \textbf{0.2857} & \underline{0.2726} & 0.2571 & 0.2477 & 0.2715 & 0.2475 \\
\midrule
\multicolumn{8}{l}{\textbf{Onion-Dedup-5000}} \\
Wav2Vec2 & 0.2407 & \textbf{0.2437} & \underline{0.2420} & 0.2299 & 0.2241 & 0.2254 & 0.2148 \\
Music2Vec & \underline{0.2614} & 0.2431 & \textbf{0.2696} & 0.2310 & 0.2512 & 0.2357 & 0.2368 \\
HuBERT & \underline{0.2409} & 0.2305 & \textbf{0.2654} & 0.2298 & 0.2263 & 0.2270 & 0.2186 \\
MERT & \underline{0.2569} & 0.2437 & \textbf{0.2674} & 0.2244 & 0.2401 & 0.2340 & 0.2171 \\
CLAP-G & \underline{0.2650} & 0.2527 & \textbf{0.2715} & 0.2442 & 0.2493 & 0.2397 & 0.2334 \\
CLAP-Music & \underline{0.2620} & 0.2533 & \textbf{0.2711} & 0.2402 & 0.2455 & 0.2492 & 0.2316 \\
\bottomrule
\end{tabular*}
\end{table}

\section{Results and Discussion}
We address four questions:
\begin{itemize}
    \item \textbf{RQ1}: How does the recommendation interface shape encoder rankings and gains from audio input?
    \item \textbf{RQ2}: How do RVQ width and depth affect ranking under the evaluated training and decoding budget?
    \item \textbf{RQ3}: Does retaining more RVQ layers improve ranking and preserve audio geometry?
    \item \textbf{RQ4}: Does stronger ranking imply a larger benefit from the original item--SID assignment?
\end{itemize}

\begin{figure*}[!t]
\centering
\includegraphics[width=\textwidth]{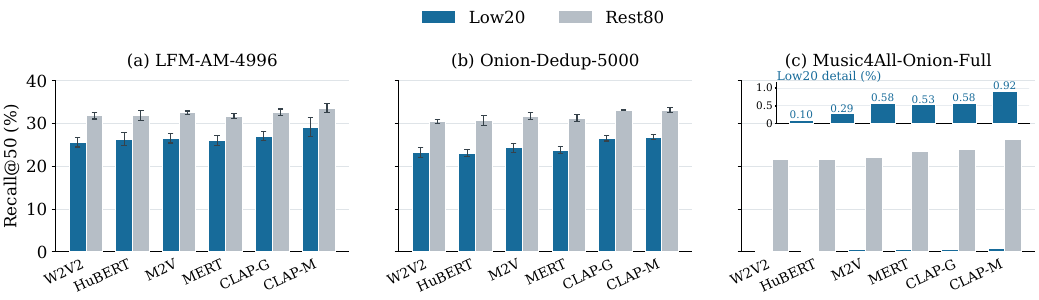}
\caption{TIGER test Recall@50 on Low20 and Rest80 tracks, using the runs in Tables~\ref{tab:main-results} and~\ref{tab:full-catalog-results}. Bars show means and error bars indicate sample standard deviations across training seeds for the two smaller catalogs; Full uses seed 2020. The inset enlarges Full Low20 values. W2V2, M2V, and CLAP-M denote Wav2Vec2, Music2Vec, and CLAP-Music.}
\label{fig:lowpop}
\Description{Three grouped bar charts compare six audio encoders on LFM-AM-4996, Onion-Dedup-5000, and Music4All-Onion-Full. Low20 Recall is lower than Rest80 for every encoder. All main panels share a zero-to-forty-percent axis; an inset shows Full Low20 values below one percent.}
\end{figure*}

\subsection{Encoder Rankings and Audio Gains (RQ1)}
\begingroup\emergencystretch=2em\relax
Table~\ref{tab:main-results} shows that encoder rankings depend on the recommendation interface. CLAP-Music ranks first in 17 of 18 dataset--recommender--metric comparisons, while CLAP-G ranks second in 14. SASRec has the highest absolute performance in all 36 dataset--encoder--metric comparisons. On the full catalog, CLAP-Music also ranks first across all three metrics within KNN and TIGER (Appendix~\ref{app:full-catalog-results}).

\paragraph{Gains over content-free inputs}
Relative encoder variation is largest under KNN, smaller under TIGER, and smallest under SASRec. Every audio encoder exceeds its dimension-matched random KNN control in mean NDCG@50 across all three catalogs. On the smaller catalogs, however, SASRec's paired audio gains are only $0.0004$--$0.0029$ over random features and $0.0002$--$0.0032$ over ID-only initialization (Table~\ref{tab:id-only}). For CLAP-Music, the ID-only gains are $0.0011$ and $0.0008$. Even with random inputs, SASRec exceeds TIGER by $0.0890$--$0.1019$ NDCG@50.

TIGER obtains larger gains from CLAP-Music: $0.0220$ and $0.0203$ NDCG@50 over matched random inputs, with positive differences in all five seed pairs on each dataset. CLAP-G also improves in every pair, whereas the 768-dimensional encoders have smaller and less consistent gains. These comparisons concern complete representation--tokenizer--generator pipelines, including input-specific tokenizers.

At seed 2020 on Full, audio-initialized SASRec falls slightly below its matched random controls for all six encoders, whereas CLAP-Music TIGER reaches $0.1696$ NDCG@50 versus $0.0059$ for random 512-dimensional inputs (Table~\ref{tab:full-catalog-results}). Thus, absolute ranking and gains associated with audio input should be evaluated separately within each interface.

\paragraph{Encoder profiles}
Among the evaluated models, audio--text-aligned and music-pretrained encoders generally perform better under KNN and TIGER; music-pretrained models win 37 of 54 paired comparisons (Music2Vec versus Wav2Vec2, MERT versus HuBERT, and CLAP-Music versus CLAP-G). These are descriptive comparisons among pretrained models, not isolated effects of training objective or domain. Their advantages shrink under SASRec alongside its modest audio gains.

\paragraph{Low-popularity tracks}
We partition tracks by distinct training listeners into the bottom 20\% (Low20) and Rest80, breaking ties by track ID, and group test queries by target track while retaining full-catalog ranking. Figure~\ref{fig:lowpop} shows lower Recall@50 for Low20 across all encoders. The gap is 4.5--6.2 percentage points on LFM-AM-4996 and 6.3--7.7 on Onion-Dedup-5000; Full Low20 reaches only $0.10\%$--$0.92\%$, versus $21.65\%$--$26.26\%$ for Rest80. Low20 still has 211--272 and 197--244 training listeners in the smaller catalogs, respectively, and 1--110 in Full, so this is a popularity analysis rather than a cold-start evaluation.

\par\endgroup

\begin{figure*}[!t]
\centering
\definecolor{w2vblue}{HTML}{0072B2}
\definecolor{hubertorange}{HTML}{E69F00}
\definecolor{m2vgreen}{HTML}{009E73}
\definecolor{mertpurple}{HTML}{CC79A7}
\definecolor{clapvermillion}{HTML}{D55E00}
\definecolor{clapmusicblue}{HTML}{56B4E9}
\begin{tikzpicture}
\begin{groupplot}[
group style={group size=5 by 2,horizontal sep=0.35cm,vertical sep=0.95cm},
scale only axis,width=0.167\textwidth,height=0.10\textwidth,
ymin=0.00,ymax=0.30,ytick={0.00,0.10,0.20,0.30},scaled y ticks=false,
yticklabel style={font=\scriptsize,/pgf/number format/fixed,/pgf/number format/precision=1},
xticklabel style={font=\scriptsize},label style={font=\footnotesize},title style={font=\footnotesize},
grid=major,major grid style={gray!35,densely dashed},axis line style={gray!70},
tick align=outside,tick style={gray!70},unbounded coords=jump,
every axis plot/.append style={line width=0.65pt,mark size=1.45pt}
]
\nextgroupplot[symbolic x coords={64,256,1024,4096},xtick={64,256,1024,4096},title={Width ($D=3$)},ylabel={LFM-AM-4996\\NDCG@50 $\uparrow$},ylabel style={align=center}]
\addplot[w2vblue,solid,mark=*] coordinates {(64,0.2273) (256,0.2367) (1024,0.2384) (4096,0.2351)};
\addplot[hubertorange,solid,mark=square*] coordinates {(64,0.2235) (256,0.2297) (1024,0.2397) (4096,0.2342)};
\addplot[m2vgreen,solid,mark=triangle*] coordinates {(64,0.2355) (256,0.2402) (1024,0.2427) (4096,0.2385)};
\addplot[mertpurple,solid,mark=diamond*] coordinates {(64,0.2385) (256,0.2304) (1024,0.2391) (4096,0.2292)};
\addplot[clapvermillion,dashed,mark=triangle*,mark options={rotate=180}] coordinates {(64,0.2401) (256,0.2407) (1024,0.2398) (4096,0.2433)};
\addplot[clapmusicblue,dashed,mark=+] coordinates {(64,0.2451) (256,0.2439) (1024,0.2367) (4096,0.2476)};
\nextgroupplot[symbolic x coords={1,3,6,12},xtick={1,3,6,12},title={Depth ($W=64$)},yticklabels={}]
\addplot[w2vblue,solid,mark=*] coordinates {(1,0.2692) (3,0.2273) (6,0.2167) (12,0.1988)};
\addplot[hubertorange,solid,mark=square*] coordinates {(1,0.2720) (3,0.2235) (6,0.2041) (12,0.1896)};
\addplot[m2vgreen,solid,mark=triangle*] coordinates {(1,0.2618) (3,0.2355) (6,0.2311) (12,0.2312)};
\addplot[mertpurple,solid,mark=diamond*] coordinates {(1,0.2588) (3,0.2385) (6,0.2143) (12,0.1958)};
\addplot[clapvermillion,dashed,mark=triangle*,mark options={rotate=180}] coordinates {(1,0.2613) (3,0.2401) (6,0.2223) (12,0.2142)};
\addplot[clapmusicblue,dashed,mark=+] coordinates {(1,0.2623) (3,0.2451) (6,0.2216) (12,0.0139)};
\nextgroupplot[symbolic x coords={1,3,6,12},xtick={1,3,6,12},title={Depth ($W=256$)},yticklabels={},legend columns=6,legend style={at={(axis description cs:0.5,1.33)},anchor=south,draw=none,font=\scriptsize,column sep=0.35em}]
\addplot[w2vblue,solid,mark=*] coordinates {(1,0.2544) (3,0.2367) (6,0.2351) (12,0.2240)};
\addlegendentry{Wav2Vec2}
\addplot[hubertorange,solid,mark=square*] coordinates {(1,0.2692) (3,0.2297) (6,0.2227) (12,0.1909)};
\addlegendentry{HuBERT}
\addplot[m2vgreen,solid,mark=triangle*] coordinates {(1,0.2596) (3,0.2402) (6,0.2419) (12,0.2293)};
\addlegendentry{Music2Vec}
\addplot[mertpurple,solid,mark=diamond*] coordinates {(1,0.2479) (3,0.2304) (6,0.2266) (12,0.2181)};
\addlegendentry{MERT}
\addplot[clapvermillion,dashed,mark=triangle*,mark options={rotate=180}] coordinates {(1,0.2558) (3,0.2407) (6,0.0105) (12,0.1969)};
\addlegendentry{CLAP-G}
\addplot[clapmusicblue,dashed,mark=+] coordinates {(1,0.2567) (3,0.2439) (6,0.2428) (12,0.0112)};
\addlegendentry{CLAP-Music}
\nextgroupplot[symbolic x coords={1,3,6,12},xtick={1,3,6,12},title={Depth ($W=1024$)},yticklabels={}]
\addplot[w2vblue,solid,mark=*] coordinates {(1,0.2557) (3,0.2384) (6,0.2458) (12,0.2307)};
\addplot[hubertorange,solid,mark=square*] coordinates {(1,0.2669) (3,0.2397) (6,0.2442) (12,0.2234)};
\addplot[m2vgreen,solid,mark=triangle*] coordinates {(1,0.2728) (3,0.2427) (6,0.2379) (12,0.2369)};
\addplot[mertpurple,solid,mark=diamond*] coordinates {(1,0.2585) (3,0.2391) (6,0.2398) (12,0.2402)};
\addplot[clapvermillion,dashed,mark=triangle*,mark options={rotate=180}] coordinates {(1,0.2538) (3,0.2398) (6,0.2512) (12,0.2388)};
\addplot[clapmusicblue,dashed,mark=+] coordinates {(1,0.2606) (3,0.2367) (6,0.2460) (12,0.2465)};
\nextgroupplot[symbolic x coords={1,3,6,12},xtick={1,3,6,12},title={Depth ($W=4096$)},yticklabels={}]
\addplot[w2vblue,solid,mark=*] coordinates {(1,0.2567) (3,0.2351) (6,0.2184) (12,0.1981)};
\addplot[hubertorange,solid,mark=square*] coordinates {(1,0.2621) (3,0.2342) (6,0.2219) (12,0.2064)};
\addplot[m2vgreen,solid,mark=triangle*] coordinates {(1,0.2691) (3,0.2385) (6,0.2398) (12,0.2257)};
\addplot[mertpurple,solid,mark=diamond*] coordinates {(1,0.2637) (3,0.2292) (6,0.2191) (12,0.2019)};
\addplot[clapvermillion,dashed,mark=triangle*,mark options={rotate=180}] coordinates {(1,0.2527) (3,0.2433) (6,0.2305) (12,0.2185)};
\addplot[clapmusicblue,dashed,mark=+] coordinates {(1,0.2588) (3,0.2476) (6,0.0074) (12,0.0042)};
\nextgroupplot[symbolic x coords={64,256,1024,4096},xtick={64,256,1024,4096},xlabel={Width $W$},ylabel={Onion-Dedup-5000\\NDCG@50 $\uparrow$},ylabel style={align=center}]
\addplot[w2vblue,solid,mark=*] coordinates {(64,0.2210) (256,0.2212) (1024,0.2258) (4096,0.2161)};
\addplot[hubertorange,solid,mark=square*] coordinates {(64,0.2101) (256,0.2270) (1024,0.2310) (4096,0.2280)};
\addplot[m2vgreen,solid,mark=triangle*] coordinates {(64,0.2233) (256,0.2280) (1024,0.2243) (4096,0.2254)};
\addplot[mertpurple,solid,mark=diamond*] coordinates {(64,0.2244) (256,0.2231) (1024,0.2266) (4096,0.2229)};
\addplot[clapvermillion,dashed,mark=triangle*,mark options={rotate=180}] coordinates {(64,0.2269) (256,0.2410) (1024,0.2376) (4096,0.2391)};
\addplot[clapmusicblue,dashed,mark=+] coordinates {(64,0.2300) (256,0.2334) (1024,0.2244) (4096,0.2369)};
\nextgroupplot[symbolic x coords={1,3,6,12},xtick={1,3,6,12},xlabel={Depth $D$},yticklabels={}]
\addplot[w2vblue,solid,mark=*] coordinates {(1,0.2572) (3,0.2210) (6,0.2159) (12,0.2067)};
\addplot[hubertorange,solid,mark=square*] coordinates {(1,0.2452) (3,0.2101) (6,0.1966) (12,0.1929)};
\addplot[m2vgreen,solid,mark=triangle*] coordinates {(1,0.2474) (3,0.2233) (6,0.2261) (12,0.2150)};
\addplot[mertpurple,solid,mark=diamond*] coordinates {(1,0.2464) (3,0.2244) (6,0.2027) (12,0.1771)};
\addplot[clapvermillion,dashed,mark=triangle*,mark options={rotate=180}] coordinates {(1,0.2561) (3,0.2269) (6,0.2086) (12,0.2055)};
\addplot[clapmusicblue,dashed,mark=+] coordinates {(1,0.2490) (3,0.2300) (6,0.2297) (12,0.2126)};
\nextgroupplot[symbolic x coords={1,3,6,12},xtick={1,3,6,12},xlabel={Depth $D$},yticklabels={}]
\addplot[w2vblue,solid,mark=*] coordinates {(1,0.2501) (3,0.2212) (6,0.2226) (12,0.1953)};
\addplot[hubertorange,solid,mark=square*] coordinates {(1,0.2547) (3,0.2270) (6,0.1969) (12,0.2021)};
\addplot[m2vgreen,solid,mark=triangle*] coordinates {(1,0.2445) (3,0.2280) (6,0.2231) (12,0.2093)};
\addplot[mertpurple,solid,mark=diamond*] coordinates {(1,0.2512) (3,0.2231) (6,0.2082) (12,0.1868)};
\addplot[clapvermillion,dashed,mark=triangle*,mark options={rotate=180}] coordinates {(1,0.2504) (3,0.2410) (6,0.2293) (12,0.2078)};
\addplot[clapmusicblue,dashed,mark=+] coordinates {(1,0.2591) (3,0.2334) (6,0.2389) (12,0.2183)};
\nextgroupplot[symbolic x coords={1,3,6,12},xtick={1,3,6,12},xlabel={Depth $D$},yticklabels={}]
\addplot[w2vblue,solid,mark=*] coordinates {(1,0.2474) (3,0.2258) (6,0.2313) (12,0.2120)};
\addplot[hubertorange,solid,mark=square*] coordinates {(1,0.2399) (3,0.2310) (6,0.2245) (12,0.2018)};
\addplot[m2vgreen,solid,mark=triangle*] coordinates {(1,0.2499) (3,0.2243) (6,0.2368) (12,0.2240)};
\addplot[mertpurple,solid,mark=diamond*] coordinates {(1,0.2498) (3,0.2266) (6,0.2385) (12,0.2206)};
\addplot[clapvermillion,dashed,mark=triangle*,mark options={rotate=180}] coordinates {(1,0.2526) (3,0.2376) (6,0.2348) (12,0.2363)};
\addplot[clapmusicblue,dashed,mark=+] coordinates {(1,0.2646) (3,0.2244) (6,0.2412) (12,0.2416)};
\nextgroupplot[symbolic x coords={1,3,6,12},xtick={1,3,6,12},xlabel={Depth $D$},yticklabels={}]
\addplot[w2vblue,solid,mark=*] coordinates {(1,0.2450) (3,0.2161) (6,0.2089) (12,0.1890)};
\addplot[hubertorange,solid,mark=square*] coordinates {(1,0.2528) (3,0.2280) (6,0.2089) (12,0.2007)};
\addplot[m2vgreen,solid,mark=triangle*] coordinates {(1,0.2541) (3,0.2254) (6,0.2201) (12,0.2110)};
\addplot[mertpurple,solid,mark=diamond*] coordinates {(1,0.2520) (3,0.2229) (6,0.2078) (12,0.1927)};
\addplot[clapvermillion,dashed,mark=triangle*,mark options={rotate=180}] coordinates {(1,0.2512) (3,0.2391) (6,0.2307) (12,0.2044)};
\addplot[clapmusicblue,dashed,mark=+] coordinates {(1,0.2599) (3,0.2369) (6,0.2314) (12,0.2217)};
\end{groupplot}
\end{tikzpicture}
\caption{Semantic-ID capacity under width ($D=3$) and depth ($W\in\{64,256,1024,4096\}$) sweeps on LFM-AM-4996 (top) and Onion-Dedup-5000 (bottom). Values are epoch-10 validation scores with seed 2020; complete results are in Table~\ref{tab:depth-sweeps}.}
\label{fig:semantic-id-capacity}
\Description{Ten panels show NDCG at 50 for a width sweep and depth sweeps at widths 64, 256, 1024, and 4096 on two datasets. Curves distinguish six audio encoders.}
\end{figure*}

\begin{figure*}[!t]
\centering
\includegraphics[width=0.75\textwidth]{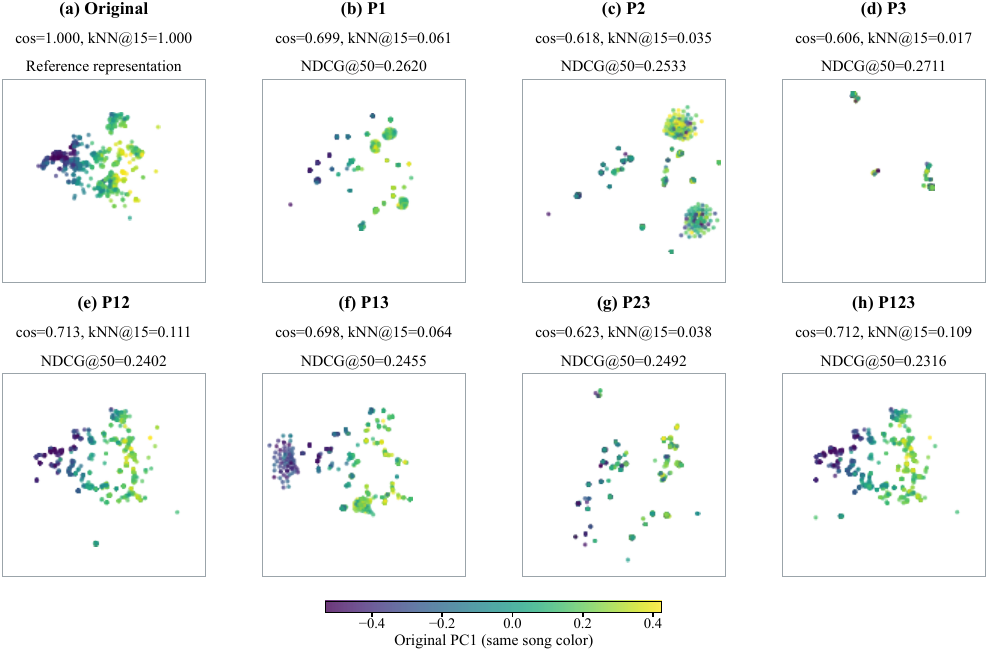}
\caption{Reconstructed audio geometry for all seven retained-layer subsets of a representative reference tokenizer (Onion-Dedup-5000, CLAP-Music, $W=256$, $D=3$). Each reconstruction decodes the sum of the selected codewords with the fixed tokenizer decoder. All eight panels use the same 1,000 sampled tracks (seed 2020), shared PCA preprocessing, and one joint t-SNE fit~\cite{vandermaaten2008tsne}. Colors follow the first principal component of the original embeddings. Cosine similarity and kNN@15 overlap are computed in the original high-dimensional space on these tracks; NDCG@50 is the corresponding epoch-10 validation result from Table~\ref{tab:prefix}.}
\label{fig:prefix-cumulative-tsne}
\Description{Eight t-SNE panels show the original audio embeddings and reconstructions from P1, P2, P3, P12, P13, P23, and P123 for the same tracks. Shared colors track items across panels. Annotations report high-dimensional cosine similarity, nearest-neighbor overlap, and validation NDCG at 50. P12 and P123 best preserve the original geometry, while P3 has the highest NDCG.}
\end{figure*}

\subsection{Semantic-ID Width and Depth (RQ2)}
Figure~\ref{fig:semantic-id-capacity} shows how codebook width and depth
affect generative recommendation (full results in Appendix~\ref{app:semantic-id-results}).
At $D=3$, larger codebooks ($W=1024$ or $4096$) achieve the best result in 27 of 36
embedding--dataset--metric comparisons with seed 2020 and in 31 of 36
comparisons using the five-seed means (Table~\ref{tab:width-multiseed}).

Under the fixed epoch-10 training and decoding protocol, $D=1$ achieves the best performance in all 36 comparisons at each of the four codebook widths. Relative to $D=1$, deeper RVQ configurations reduce performance, although the changes among deeper configurations are not monotonic. Severe degradation occurs on LFM-AM-4996 for CLAP-Music at $W=4096$, $D=6$ and $12$, and at $W=64$ or $256$, $D=12$, as well as for CLAP-G at $W=256$, $D=6$. These results favor $D=1$ within each tested width and larger codebooks in the separate $D=3$ width sweep.
We next examine whether these ranking patterns correspond to codebook usage.
Layer-wise codebook usage changes with width (Figure~\ref{fig:codebook-width-usage}).
At $D=3$, the active-codeword fraction averaged over encoders and datasets
increases from the first to the third layer for $W=256$ and $1024$, whereas
the third-layer fraction averages $1.48\%$ at $W=4096$. Across the width and depth sweeps (24 configurations per sweep), raw Pearson correlations between mean layer usage and validation NDCG@50 range from $-0.32$ to $0.06$ on LFM-AM-4996 and from $-0.32$ to $0.66$ on Onion-Dedup-5000, showing that higher usage does not consistently correspond to better ranking.

\subsection{Retained Layers and Audio Geometry (RQ3)}

Table~\ref{tab:prefix} reports retained-layer NDCG@50; Recall@50 and MRR@50
are reported in Appendix~\ref{app:semantic-id-results}. Single-layer subsets achieve the best performance
in all 36 embedding--dataset--metric comparisons (P1: 5, P2: 8, and P3: 23). In contrast, P123 does
not achieve the best result in any setting and is worse than P12 in 29 out of
36 comparisons.

These results show that a Semantic ID formed from one retained RVQ layer and its suffix can support effective ranking. Combining all three layers provides no benefit over the best single-layer subset in this comparison. The assignment controls below examine how much of this performance depends on the original item--SID association.

Figure~\ref{fig:prefix-cumulative-tsne} relates these ranking results to reconstructed
audio geometry for CLAP-Music on Onion-Dedup-5000. P12 and P123 best preserve
the original embeddings, with mean cosine similarity of 0.7129 and 0.7123
and kNN@15 overlap of 0.1107 and 0.1095, respectively. Yet their NDCG@50
scores are 0.2402 and 0.2316, below all three single-layer subsets.
P3 achieves the highest NDCG@50 (0.2711), despite having the lowest cosine
similarity (0.6056) and neighborhood overlap (0.0173).

Together, Table~\ref{tab:prefix} and Figure~\ref{fig:prefix-cumulative-tsne} show that
faithful reconstruction of the pretrained audio geometry does not necessarily
yield better recommendation. Combining layers can improve reconstruction without improving ranking. This is consistent with RQ2:
increasing Semantic-ID complexity need not improve recommendation performance.
Layer-wise usage and perplexity for these fixed tokenizers are reported in Table~\ref{tab:prefix-codebook-usage}.

\subsection{Ranking and Assignment Gains (RQ4)}
Table~\ref{tab:sid-assignment} evaluates P3 and P123 with the CLAP-Music reference tokenizer under original and shuffled assignments. P3 ranks higher under both assignments on both datasets. On complete four-condition blocks, its NDCG@50 advantage over P123 is $0.0243$ on LFM-AM-4996 and $0.0383$ on Onion-Dedup-5000 under the original assignment, increasing to $0.0422$ and $0.0718$ after shuffling.

The original assignment benefits P123 more. Its paired gains are $0.0254$ and $0.0358$, compared with $0.0076$ and $0.0023$ for P3; the corresponding interactions are $-0.0179$ and $-0.0335$. The interaction is negative in every complete seed block, with the same direction for Recall@50 and MRR@50. P3's ranking advantage persists after shuffling, while its gain from the original assignment is smaller. The comparison evaluates complete identifiers, including retained codes, collision suffixes, and active prediction positions. Shuffling also changes code frequencies in interaction sequences, so the gains measure the benefit of the complete item--SID assignment. The results favor evaluating Semantic IDs by both ranking quality and assignment sensitivity under the chosen training budget.

\begin{table}[!t]
\centering\footnotesize
\setlength{\tabcolsep}{1.2pt}
\renewcommand{\arraystretch}{1.0}
\caption{CLAP-Music assignment controls with the retained-layer reference tokenizer (epoch-10 test, last-100 exclusion). Top: mean $\pm$ sample SD and run count $n$. Bottom: mean NDCG@50 contrasts over $n$ complete four-condition seed blocks: $G_S=\mathrm{original}-\mathrm{shuffled}$, $I=G_{\mathrm{P3}}-G_{\mathrm{P123}}$, and $\Delta_{\mathrm{orig/shuf}}=\mathrm{P3}-\mathrm{P123}$ within each assignment.}
\label{tab:sid-assignment}
\begin{tabular*}{\columnwidth}{@{\extracolsep{\fill}}llcrrr@{}}
\toprule
Subset & Assignment & $n$ & R@50 & N@50 & MRR@50 \\
\midrule

\multicolumn{6}{@{}l}{\textbf{LFM-AM-4996}} \\
P3 & Original & 5 & $0.3851\pm0.0034$ & $0.2395\pm0.0026$ & $0.2021\pm0.0025$ \\
P3 & Shuffled & 5 & $0.3731\pm0.0035$ & $0.2319\pm0.0011$ & $0.1959\pm0.0006$ \\
P123 & Original & 5 & $0.3375\pm0.0036$ & $0.2152\pm0.0036$ & $0.1841\pm0.0038$ \\
P123 & Shuffled & 5 & $0.2910\pm0.0054$ & $0.1897\pm0.0027$ & $0.1643\pm0.0022$ \\
\midrule
\multicolumn{6}{@{}l}{\textbf{Onion-Dedup-5000}} \\
P3 & Original & 5 & $0.3827\pm0.0050$ & $0.2397\pm0.0029$ & $0.2022\pm0.0024$ \\
P3 & Shuffled & 5 & $0.3790\pm0.0046$ & $0.2374\pm0.0019$ & $0.2002\pm0.0015$ \\
P123 & Original & 5 & $0.3304\pm0.0029$ & $0.2014\pm0.0027$ & $0.1692\pm0.0028$ \\
P123 & Shuffled & 5 & $0.2670\pm0.0076$ & $0.1656\pm0.0080$ & $0.1406\pm0.0080$ \\
\bottomrule
\end{tabular*}
\par\smallskip
\begin{tabular*}{\columnwidth}{@{\extracolsep{\fill}}lrr@{}}
\toprule
Contrast & LFM-AM-4996 & Onion-Dedup-5000 \\
\midrule
$n$ & 5 & 5 \\
$G_{\mathrm{P3}}$ & $+0.0076$ & $+0.0023$ \\
$G_{\mathrm{P123}}$ & $+0.0254$ & $+0.0358$ \\
$\Delta_{\mathrm{orig}}$ & $+0.0243$ & $+0.0383$ \\
$\Delta_{\mathrm{shuf}}$ & $+0.0422$ & $+0.0718$ \\
$I$ & $-0.0179$ & $-0.0335$ \\
\bottomrule
\end{tabular*}
\end{table}

\section{Conclusion}
Audio representation utility depends on the recommendation interface. CLAP-Music performs well under KNN and TIGER, whereas SASRec narrows encoder differences and remains competitive with random or ID-only inputs. At epoch 10, depth one ranks best at each tested width; larger codebooks generally help at depth three. Selected single-layer identifiers outperform the full three-layer representation in every evaluated comparison. Better audio reconstruction or larger gains from the original item--SID mapping do not necessarily yield better ranking. Audio representations and Semantic IDs should therefore be assessed by downstream ranking alongside matched content-free and assignment controls.

\bibliographystyle{ACM-Reference-Format}
\bibliography{v25_references}
\ifdefined\VtwentyfourReferenceReserve
\par\noindent\fbox{\parbox{\dimexpr\columnwidth-2\fboxsep-2\fboxrule\relax}{\centering Reference-space reserve: ten additional entries}}\par
\vspace*{480pt}
\fi

\appendix
\makeatletter
\setlength{\@fptop}{0pt}
\setlength{\@dblfptop}{0pt}
\setlength{\@fpsep}{12pt}
\setlength{\@dblfpsep}{12pt}
\makeatother

\begin{table*}[!t]
\captionsetup{skip=4pt}
\caption{Width sweep at $D=3$ on LFM-AM-4996 and Onion-Dedup-5000. Results are epoch-10 validation mean $\pm$ sample standard deviation over generator seeds 2020--2024, with a fixed tokenizer per condition. \textbf{Bold} and \underline{underlined} values denote the best and second-best widths within each encoder, dataset, and metric, ranked using unrounded means. N denotes NDCG.}
\label{tab:width-multiseed}
\centering
\scriptsize
\setlength{\tabcolsep}{2.5pt}
\renewcommand{\arraystretch}{0.95}
\begin{tabular}{l|r|ccc|ccc}
\toprule
\multirow{3}{*}{\textbf{Embedding}} & & \multicolumn{6}{c}{\textbf{Width Sweep} ($D=3$)} \\
\cmidrule(lr){3-8}
& & \multicolumn{3}{c|}{\textbf{LFM-AM-4996}} & \multicolumn{3}{c}{\textbf{Onion-Dedup-5000}} \\
\cmidrule(lr){3-5}\cmidrule(lr){6-8}
& $W$ & R@50 & N@50 & MRR@50 & R@50 & N@50 & MRR@50 \\
\midrule
\multirow{4}{*}{Wav2Vec2} & 64 & $0.3548\pm0.0059$ & $0.2337\pm0.0055$ & $0.2027\pm0.0055$ & $0.3515\pm0.0026$ & $\underline{0.2228\pm0.0020}$ & $\underline{0.1904\pm0.0019}$ \\
 & 256 & $0.3569\pm0.0074$ & $0.2330\pm0.0056$ & $0.2012\pm0.0051$ & $\underline{0.3517\pm0.0036}$ & $0.2209\pm0.0017$ & $0.1878\pm0.0017$ \\
 & 1024 & $\mathbf{0.3625\pm0.0031}$ & $\mathbf{0.2394\pm0.0011}$ & $\mathbf{0.2073\pm0.0012}$ & $\mathbf{0.3524\pm0.0028}$ & $\mathbf{0.2247\pm0.0019}$ & $\mathbf{0.1922\pm0.0017}$ \\
 & 4096 & $\underline{0.3579\pm0.0024}$ & $\underline{0.2348\pm0.0012}$ & $\underline{0.2031\pm0.0014}$ & $0.3421\pm0.0056$ & $0.2189\pm0.0031$ & $0.1879\pm0.0025$ \\
\hline
\multirow{4}{*}{Music2Vec} & 64 & $0.3582\pm0.0038$ & $0.2361\pm0.0024$ & $0.2047\pm0.0024$ & $0.3542\pm0.0041$ & $0.2244\pm0.0019$ & $0.1917\pm0.0014$ \\
 & 256 & $0.3648\pm0.0027$ & $\mathbf{0.2412\pm0.0010}$ & $\mathbf{0.2092\pm0.0015}$ & $\mathbf{0.3626\pm0.0057}$ & $\underline{0.2296\pm0.0028}$ & $\underline{0.1957\pm0.0026}$ \\
 & 1024 & $\underline{0.3669\pm0.0053}$ & $0.2394\pm0.0046$ & $0.2064\pm0.0045$ & $\underline{0.3616\pm0.0075}$ & $\mathbf{0.2299\pm0.0044}$ & $\mathbf{0.1965\pm0.0038}$ \\
 & 4096 & $\mathbf{0.3705\pm0.0011}$ & $\underline{0.2410\pm0.0027}$ & $\underline{0.2073\pm0.0031}$ & $0.3510\pm0.0024$ & $0.2275\pm0.0013$ & $0.1956\pm0.0014$ \\
\hline
\multirow{4}{*}{HuBERT} & 64 & $0.3388\pm0.0054$ & $0.2256\pm0.0044$ & $0.1968\pm0.0041$ & $0.3303\pm0.0085$ & $0.2116\pm0.0056$ & $0.1819\pm0.0048$ \\
 & 256 & $0.3423\pm0.0039$ & $0.2281\pm0.0021$ & $0.1988\pm0.0016$ & $0.3426\pm0.0077$ & $0.2187\pm0.0056$ & $0.1876\pm0.0053$ \\
 & 1024 & $\mathbf{0.3666\pm0.0051}$ & $\mathbf{0.2408\pm0.0038}$ & $\mathbf{0.2083\pm0.0035}$ & $\mathbf{0.3602\pm0.0023}$ & $\mathbf{0.2292\pm0.0011}$ & $\underline{0.1957\pm0.0011}$ \\
 & 4096 & $\underline{0.3551\pm0.0012}$ & $\underline{0.2348\pm0.0009}$ & $\underline{0.2038\pm0.0009}$ & $\underline{0.3506\pm0.0040}$ & $\underline{0.2280\pm0.0021}$ & $\mathbf{0.1962\pm0.0017}$ \\
\hline
\multirow{4}{*}{MERT} & 64 & $\mathbf{0.3594\pm0.0080}$ & $\underline{0.2373\pm0.0059}$ & $\underline{0.2061\pm0.0053}$ & $0.3519\pm0.0068$ & $0.2218\pm0.0041$ & $\underline{0.1893\pm0.0034}$ \\
 & 256 & $0.3481\pm0.0045$ & $0.2320\pm0.0038$ & $0.2021\pm0.0037$ & $0.3388\pm0.0131$ & $0.2156\pm0.0101$ & $0.1846\pm0.0092$ \\
 & 1024 & $\underline{0.3565\pm0.0045}$ & $\mathbf{0.2381\pm0.0018}$ & $\mathbf{0.2071\pm0.0013}$ & $\mathbf{0.3600\pm0.0084}$ & $\mathbf{0.2239\pm0.0078}$ & $\mathbf{0.1894\pm0.0078}$ \\
 & 4096 & $0.3540\pm0.0048$ & $0.2325\pm0.0028$ & $0.2013\pm0.0025$ & $\underline{0.3572\pm0.0019}$ & $\underline{0.2219\pm0.0020}$ & $0.1877\pm0.0021$ \\
\hline
\multirow{4}{*}{CLAP-G} & 64 & $0.3623\pm0.0011$ & $0.2410\pm0.0013$ & $0.2096\pm0.0017$ & $0.3669\pm0.0059$ & $0.2303\pm0.0041$ & $0.1959\pm0.0036$ \\
 & 256 & $\mathbf{0.3648\pm0.0063}$ & $\underline{0.2424\pm0.0027}$ & $\underline{0.2104\pm0.0021}$ & $\underline{0.3719\pm0.0066}$ & $\underline{0.2364\pm0.0036}$ & $\underline{0.2020\pm0.0029}$ \\
 & 1024 & $\underline{0.3639\pm0.0018}$ & $0.2398\pm0.0014$ & $0.2074\pm0.0014$ & $0.3651\pm0.0081$ & $0.2322\pm0.0051$ & $0.1980\pm0.0044$ \\
 & 4096 & $0.3628\pm0.0021$ & $\mathbf{0.2444\pm0.0011}$ & $\mathbf{0.2131\pm0.0015}$ & $\mathbf{0.3720\pm0.0033}$ & $\mathbf{0.2387\pm0.0024}$ & $\mathbf{0.2041\pm0.0024}$ \\
\hline
\multirow{4}{*}{CLAP-Music} & 64 & $0.3697\pm0.0082$ & $0.2410\pm0.0065$ & $0.2080\pm0.0062$ & $0.3665\pm0.0135$ & $0.2288\pm0.0082$ & $0.1941\pm0.0068$ \\
 & 256 & $\underline{0.3737\pm0.0026}$ & $\underline{0.2433\pm0.0020}$ & $\underline{0.2097\pm0.0020}$ & $\underline{0.3701\pm0.0037}$ & $\underline{0.2316\pm0.0036}$ & $\underline{0.1965\pm0.0037}$ \\
 & 1024 & $0.3501\pm0.0058$ & $0.2308\pm0.0048$ & $0.2002\pm0.0047$ & $0.3490\pm0.0031$ & $0.2276\pm0.0031$ & $0.1962\pm0.0031$ \\
 & 4096 & $\mathbf{0.3747\pm0.0027}$ & $\mathbf{0.2471\pm0.0035}$ & $\mathbf{0.2137\pm0.0038}$ & $\mathbf{0.3703\pm0.0081}$ & $\mathbf{0.2325\pm0.0054}$ & $\mathbf{0.1976\pm0.0048}$ \\
\bottomrule
\end{tabular}
\par\smallskip\captionsetup{type=figure}
\centering
\includegraphics[width=\textwidth]{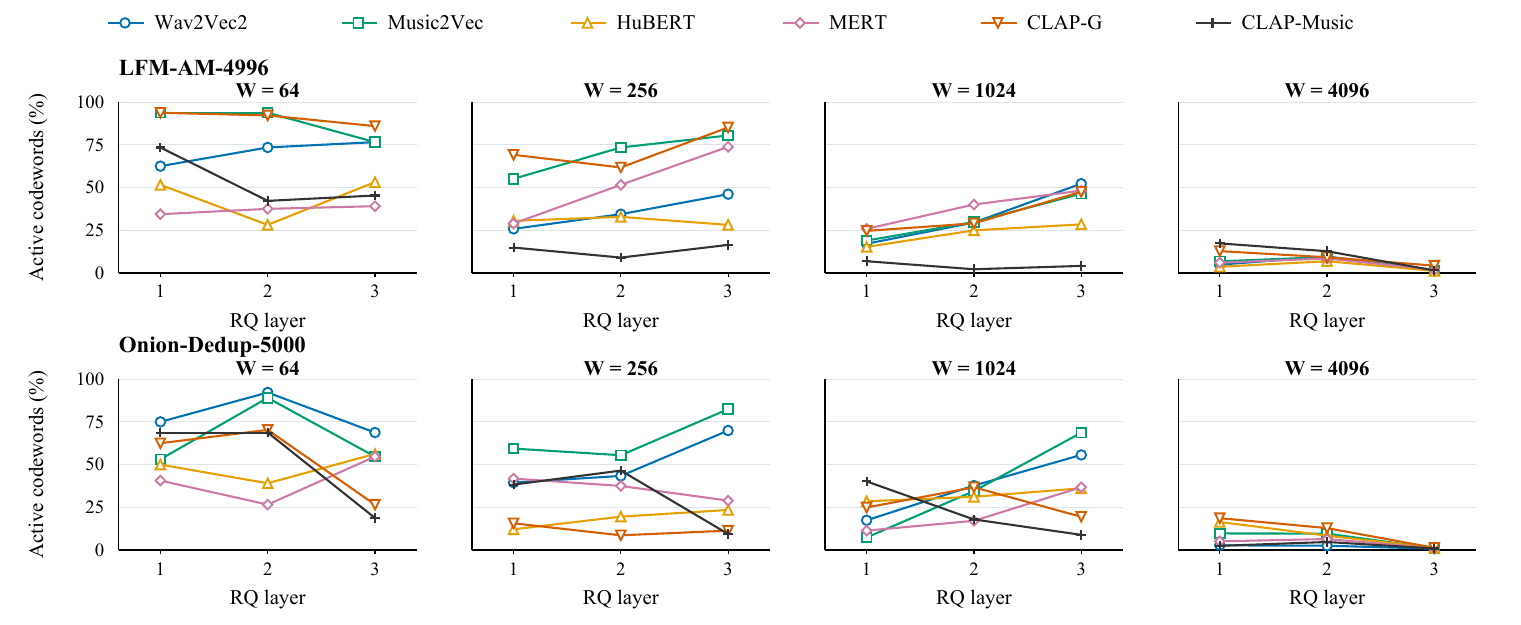}
\caption{Layer-wise active-codeword usage for the $D=3$ width sweep in Table~\ref{tab:width-multiseed}. Rows show LFM-AM-4996 and Onion-Dedup-5000; columns show codebook widths. The horizontal axis is RQ layer (1--3). Each curve represents one encoder.}
\label{fig:codebook-width-usage}
\Description{Eight line charts show active-codeword fractions across RQ layers 1, 2, and 3 at fixed total depth 3. Rows separate the two datasets and columns separate widths 64, 256, 1024, and 4096; each curve follows one encoder's tokenizer.}
\end{table*}

\begin{table*}[!t]
\captionsetup{skip=4pt}
\centering
\caption{Depth sweeps at $W\in\{64,256,1024,4096\}$. Each width is evaluated on LFM-AM-4996 and Onion-Dedup-5000. Results are epoch-10 validation scores with seed 2020.}
\label{tab:depth-sweeps}
\scriptsize
\setlength{\tabcolsep}{2.4pt}
\renewcommand{\arraystretch}{0.95}
\begin{tabular*}{\textwidth}{@{\extracolsep{\fill}}llrrr|rrr|rrr|rrr@{}}
\toprule
\multirow{3}{*}{Embedding} & \multirow{3}{*}{$D$} & \multicolumn{6}{c|}{$W=64$} & \multicolumn{6}{c}{$W=256$} \\
\cmidrule(lr){3-8}\cmidrule(lr){9-14}
& & \multicolumn{3}{c|}{LFM-AM-4996} & \multicolumn{3}{c|}{Onion-Dedup-5000} & \multicolumn{3}{c|}{LFM-AM-4996} & \multicolumn{3}{c}{Onion-Dedup-5000} \\
\cmidrule(lr){3-5}\cmidrule(lr){6-8}\cmidrule(lr){9-11}\cmidrule(lr){12-14}
& & R@50 & N@50 & MRR@50 & R@50 & N@50 & MRR@50 & R@50 & N@50 & MRR@50 & R@50 & N@50 & MRR@50 \\
\midrule
\multirow{4}{*}{Wav2Vec2} & 1 & \textbf{0.4158} & \textbf{0.2692} & \textbf{0.2311} & \textbf{0.4124} & \textbf{0.2572} & \textbf{0.2172} & \textbf{0.3844} & \textbf{0.2544} & \textbf{0.2205} & \textbf{0.4026} & \textbf{0.2501} & \textbf{0.2112} \\
 & 3 & \underline{0.3467} & \underline{0.2273} & \underline{0.1967} & \underline{0.3485} & \underline{0.2210} & \underline{0.1890} & 0.3620 & \underline{0.2367} & \underline{0.2045} & \underline{0.3564} & 0.2212 & 0.1874 \\
 & 6 & 0.3310 & 0.2167 & 0.1879 & 0.3361 & 0.2159 & 0.1857 & \underline{0.3625} & 0.2351 & 0.2025 & 0.3495 & \underline{0.2226} & \underline{0.1906} \\
 & 12 & 0.3061 & 0.1988 & 0.1718 & 0.3190 & 0.2067 & 0.1785 & 0.3462 & 0.2240 & 0.1928 & 0.3001 & 0.1953 & 0.1686 \\
\addlinespace[2pt]
\multirow{4}{*}{Music2Vec} & 1 & \textbf{0.4089} & \textbf{0.2618} & \textbf{0.2238} & \textbf{0.3868} & \textbf{0.2474} & \textbf{0.2116} & \textbf{0.3926} & \textbf{0.2596} & \textbf{0.2249} & \textbf{0.3766} & \textbf{0.2445} & \textbf{0.2105} \\
 & 3 & \underline{0.3615} & \underline{0.2355} & \underline{0.2032} & \underline{0.3525} & 0.2233 & 0.1909 & 0.3677 & 0.2402 & \underline{0.2072} & 0.3592 & \underline{0.2280} & \underline{0.1946} \\
 & 6 & 0.3570 & 0.2311 & 0.1991 & 0.3508 & \underline{0.2261} & \underline{0.1945} & \underline{0.3779} & \underline{0.2419} & 0.2069 & \underline{0.3621} & 0.2231 & 0.1880 \\
 & 12 & 0.3501 & 0.2312 & 0.2007 & 0.3392 & 0.2150 & 0.1837 & 0.3533 & 0.2293 & 0.1974 & 0.3356 & 0.2093 & 0.1773 \\
\addlinespace[2pt]
\multirow{4}{*}{HuBERT} & 1 & \textbf{0.4218} & \textbf{0.2720} & \textbf{0.2329} & \textbf{0.3933} & \textbf{0.2452} & \textbf{0.2077} & \textbf{0.4127} & \textbf{0.2692} & \textbf{0.2312} & \textbf{0.4081} & \textbf{0.2547} & \textbf{0.2155} \\
 & 3 & \underline{0.3373} & \underline{0.2235} & \underline{0.1946} & \underline{0.3270} & \underline{0.2101} & \underline{0.1810} & \underline{0.3446} & \underline{0.2297} & \underline{0.2001} & \underline{0.3493} & \underline{0.2270} & \underline{0.1960} \\
 & 6 & 0.3081 & 0.2041 & 0.1781 & 0.3086 & 0.1966 & 0.1688 & 0.3375 & 0.2227 & 0.1933 & 0.3191 & 0.1969 & 0.1664 \\
 & 12 & 0.2840 & 0.1896 & 0.1659 & 0.3050 & 0.1929 & 0.1648 & 0.2966 & 0.1909 & 0.1641 & 0.3182 & 0.2021 & 0.1730 \\
\addlinespace[2pt]
\multirow{4}{*}{MERT} & 1 & \textbf{0.3994} & \textbf{0.2588} & \textbf{0.2224} & \textbf{0.3918} & \textbf{0.2464} & \textbf{0.2093} & \textbf{0.3666} & \textbf{0.2479} & \textbf{0.2168} & \textbf{0.4019} & \textbf{0.2512} & \textbf{0.2128} \\
 & 3 & \underline{0.3631} & \underline{0.2385} & \underline{0.2066} & \underline{0.3571} & \underline{0.2244} & \underline{0.1913} & \underline{0.3483} & \underline{0.2304} & \underline{0.2002} & \underline{0.3484} & \underline{0.2231} & \underline{0.1914} \\
 & 6 & 0.3201 & 0.2143 & 0.1874 & 0.3145 & 0.2027 & 0.1749 & 0.3398 & 0.2266 & 0.1973 & 0.3322 & 0.2082 & 0.1772 \\
 & 12 & 0.2969 & 0.1958 & 0.1705 & 0.2707 & 0.1771 & 0.1538 & 0.3347 & 0.2181 & 0.1880 & 0.2985 & 0.1868 & 0.1588 \\
\addlinespace[2pt]
\multirow{4}{*}{CLAP-G} & 1 & \textbf{0.3906} & \textbf{0.2613} & \textbf{0.2276} & \textbf{0.4025} & \textbf{0.2561} & \textbf{0.2184} & \textbf{0.3821} & \textbf{0.2558} & \textbf{0.2228} & \textbf{0.3876} & \textbf{0.2504} & \textbf{0.2149} \\
 & 3 & \underline{0.3633} & \underline{0.2401} & \underline{0.2081} & \underline{0.3623} & \underline{0.2269} & \underline{0.1927} & \underline{0.3607} & \underline{0.2407} & \underline{0.2091} & \underline{0.3785} & \underline{0.2410} & \underline{0.2060} \\
 & 6 & 0.3374 & 0.2223 & 0.1929 & 0.3336 & 0.2086 & 0.1774 & 0.0362 & 0.0105 & 0.0044 & 0.3639 & 0.2293 & 0.1949 \\
 & 12 & 0.3220 & 0.2142 & 0.1866 & 0.3310 & 0.2055 & 0.1739 & 0.2969 & 0.1969 & 0.1715 & 0.3240 & 0.2078 & 0.1783 \\
\addlinespace[2pt]
\multirow{4}{*}{CLAP-Music} & 1 & \textbf{0.4038} & \textbf{0.2623} & \textbf{0.2256} & \textbf{0.3846} & \textbf{0.2490} & \textbf{0.2140} & \textbf{0.3915} & \textbf{0.2567} & \textbf{0.2215} & \textbf{0.4085} & \textbf{0.2591} & \textbf{0.2207} \\
 & 3 & \underline{0.3782} & \underline{0.2451} & \underline{0.2111} & 0.3710 & \underline{0.2300} & \underline{0.1944} & \underline{0.3739} & \underline{0.2439} & \underline{0.2105} & 0.3703 & 0.2334 & 0.1989 \\
 & 6 & 0.3330 & 0.2216 & 0.1933 & \underline{0.3715} & 0.2297 & 0.1938 & 0.3727 & 0.2428 & 0.2091 & \underline{0.3779} & \underline{0.2389} & \underline{0.2033} \\
 & 12 & 0.0432 & 0.0139 & 0.0070 & 0.3394 & 0.2126 & 0.1808 & 0.0406 & 0.0112 & 0.0045 & 0.3553 & 0.2183 & 0.1838 \\
\midrule
\addlinespace[3pt]
\multirow{3}{*}{Embedding} & \multirow{3}{*}{$D$} & \multicolumn{6}{c|}{$W=1024$} & \multicolumn{6}{c}{$W=4096$} \\
\cmidrule(lr){3-8}\cmidrule(lr){9-14}
& & \multicolumn{3}{c|}{LFM-AM-4996} & \multicolumn{3}{c|}{Onion-Dedup-5000} & \multicolumn{3}{c|}{LFM-AM-4996} & \multicolumn{3}{c}{Onion-Dedup-5000} \\
\cmidrule(lr){3-5}\cmidrule(lr){6-8}\cmidrule(lr){9-11}\cmidrule(lr){12-14}
& & R@50 & N@50 & MRR@50 & R@50 & N@50 & MRR@50 & R@50 & N@50 & MRR@50 & R@50 & N@50 & MRR@50 \\
\midrule
\multirow{4}{*}{Wav2Vec2} & 1 & \textbf{0.3900} & \textbf{0.2557} & \textbf{0.2207} & \textbf{0.3948} & \textbf{0.2474} & \textbf{0.2100} & \textbf{0.3904} & \textbf{0.2567} & \textbf{0.2215} & \textbf{0.3863} & \textbf{0.2450} & \textbf{0.2088} \\
 & 3 & 0.3647 & 0.2384 & 0.2055 & 0.3542 & 0.2258 & 0.1931 & \underline{0.3613} & \underline{0.2351} & \underline{0.2025} & \underline{0.3372} & \underline{0.2161} & \underline{0.1856} \\
 & 6 & \underline{0.3854} & \underline{0.2458} & \underline{0.2096} & \underline{0.3693} & \underline{0.2313} & \underline{0.1961} & 0.3322 & 0.2184 & 0.1895 & 0.3267 & 0.2089 & 0.1793 \\
 & 12 & 0.3577 & 0.2307 & 0.1980 & 0.3430 & 0.2120 & 0.1788 & 0.3090 & 0.1981 & 0.1703 & 0.3077 & 0.1890 & 0.1593 \\
\addlinespace[2pt]
\multirow{4}{*}{Music2Vec} & 1 & \textbf{0.4127} & \textbf{0.2728} & \textbf{0.2355} & \textbf{0.3904} & \textbf{0.2499} & \textbf{0.2135} & \textbf{0.4053} & \textbf{0.2691} & \textbf{0.2328} & \textbf{0.3957} & \textbf{0.2541} & \textbf{0.2175} \\
 & 3 & \underline{0.3714} & \underline{0.2427} & \underline{0.2098} & 0.3560 & 0.2243 & 0.1912 & \underline{0.3700} & 0.2385 & 0.2043 & \underline{0.3486} & \underline{0.2254} & \underline{0.1937} \\
 & 6 & 0.3706 & 0.2379 & 0.2037 & \underline{0.3755} & \underline{0.2368} & \underline{0.2012} & 0.3611 & \underline{0.2398} & \underline{0.2081} & 0.3467 & 0.2201 & 0.1879 \\
 & 12 & 0.3646 & 0.2369 & 0.2040 & 0.3600 & 0.2240 & 0.1894 & 0.3446 & 0.2257 & 0.1951 & 0.3276 & 0.2110 & 0.1815 \\
\addlinespace[2pt]
\multirow{4}{*}{HuBERT} & 1 & \textbf{0.4113} & \textbf{0.2669} & \textbf{0.2291} & \textbf{0.3819} & \textbf{0.2399} & \textbf{0.2035} & \textbf{0.3994} & \textbf{0.2621} & \textbf{0.2261} & \textbf{0.4023} & \textbf{0.2528} & \textbf{0.2142} \\
 & 3 & 0.3661 & 0.2397 & 0.2072 & \underline{0.3618} & \underline{0.2310} & \underline{0.1976} & \underline{0.3550} & \underline{0.2342} & \underline{0.2032} & \underline{0.3484} & \underline{0.2280} & \underline{0.1969} \\
 & 6 & \underline{0.3837} & \underline{0.2442} & \underline{0.2084} & 0.3435 & 0.2245 & 0.1937 & 0.3354 & 0.2219 & 0.1926 & 0.3266 & 0.2089 & 0.1790 \\
 & 12 & 0.3357 & 0.2234 & 0.1943 & 0.3321 & 0.2018 & 0.1690 & 0.3161 & 0.2064 & 0.1784 & 0.3098 & 0.2007 & 0.1727 \\
\addlinespace[2pt]
\multirow{4}{*}{MERT} & 1 & \textbf{0.4035} & \textbf{0.2585} & \textbf{0.2208} & \textbf{0.3997} & \textbf{0.2498} & \textbf{0.2112} & \textbf{0.4009} & \textbf{0.2637} & \textbf{0.2276} & \textbf{0.3998} & \textbf{0.2520} & \textbf{0.2138} \\
 & 3 & 0.3611 & 0.2391 & \underline{0.2074} & 0.3625 & 0.2266 & 0.1921 & \underline{0.3458} & \underline{0.2292} & \underline{0.1992} & \underline{0.3558} & \underline{0.2229} & \underline{0.1893} \\
 & 6 & 0.3655 & 0.2398 & 0.2070 & \underline{0.3808} & \underline{0.2385} & \underline{0.2023} & 0.3303 & 0.2191 & 0.1904 & 0.3291 & 0.2078 & 0.1774 \\
 & 12 & \underline{0.3791} & \underline{0.2402} & 0.2045 & 0.3552 & 0.2206 & 0.1861 & 0.3061 & 0.2019 & 0.1752 & 0.3073 & 0.1927 & 0.1641 \\
\addlinespace[2pt]
\multirow{4}{*}{CLAP-G} & 1 & \textbf{0.3794} & \textbf{0.2538} & \textbf{0.2208} & \textbf{0.3960} & \textbf{0.2526} & \textbf{0.2153} & \textbf{0.3787} & \textbf{0.2527} & \textbf{0.2198} & \textbf{0.3885} & \textbf{0.2512} & \textbf{0.2156} \\
 & 3 & 0.3655 & 0.2398 & 0.2069 & 0.3726 & \underline{0.2376} & \underline{0.2029} & \underline{0.3648} & \underline{0.2433} & \underline{0.2113} & \underline{0.3717} & \underline{0.2391} & \underline{0.2047} \\
 & 6 & \underline{0.3792} & \underline{0.2512} & \underline{0.2178} & 0.3590 & 0.2348 & 0.2023 & 0.3499 & 0.2305 & 0.1997 & 0.3563 & 0.2307 & 0.1980 \\
 & 12 & 0.3668 & 0.2388 & 0.2053 & \underline{0.3833} & 0.2363 & 0.1988 & 0.3309 & 0.2185 & 0.1896 & 0.3292 & 0.2044 & 0.1725 \\
\addlinespace[2pt]
\multirow{4}{*}{CLAP-Music} & 1 & \textbf{0.3955} & \textbf{0.2606} & \textbf{0.2253} & \textbf{0.4185} & \textbf{0.2646} & \textbf{0.2248} & \textbf{0.3991} & \textbf{0.2588} & \textbf{0.2227} & \textbf{0.4081} & \textbf{0.2599} & \textbf{0.2214} \\
 & 3 & 0.3568 & 0.2367 & 0.2058 & 0.3452 & 0.2244 & 0.1929 & \underline{0.3756} & \underline{0.2476} & \underline{0.2140} & \underline{0.3752} & \underline{0.2369} & \underline{0.2019} \\
 & 6 & \underline{0.3851} & 0.2460 & 0.2102 & 0.3828 & 0.2412 & \underline{0.2046} & 0.0260 & 0.0074 & 0.0031 & 0.3700 & 0.2314 & 0.1964 \\
 & 12 & 0.3757 & \underline{0.2465} & \underline{0.2125} & \underline{0.3884} & \underline{0.2416} & 0.2040 & 0.0159 & 0.0042 & 0.0015 & 0.3496 & 0.2217 & 0.1890 \\
\bottomrule
\end{tabular*}
\end{table*}

\section{Supplementary Semantic-ID Results}
\label{app:semantic-id-results}
Table~\ref{tab:width-multiseed} reports the five-seed width results at $D=3$.
Table~\ref{tab:depth-sweeps} reports the seed-2020 depth sweeps across all four
widths, covering 192 unique configurations; its $D=3$ rows also provide the
width-sweep results used in Figure~\ref{fig:semantic-id-capacity}.
Bold and underlined values in Table~\ref{tab:depth-sweeps} mark the best and
second-best depths within each encoder, dataset, width, and metric, ranked
using unrounded scores.
Table~\ref{tab:prefix-full} complements the retained-layer NDCG@50 results in Table~\ref{tab:prefix} with Recall@50 and MRR@50.

\section{Full-Catalog Results}\label{app:full-catalog-results}
Table~\ref{tab:full-catalog-results} evaluates all three recommenders on
56,512 Music4All-Onion tracks and the same 116,583 test users and targets,
using the histories and candidate filtering in
Section~\ref{sec:experiment-settings}. SASRec and TIGER use epoch-10
checkpoints with seed 2020; KNN is evaluated once. TIGER uses beam size 200
and 20 expansions per beam. Full uses separate frozen CLAP feature
extractions from Onion-Dedup-5000.

\section{Codebook Diagnostics}
\label{app:codebook-diagnostics}
Usage is $K_{\mathrm{active}}/W$, counting each catalog track once. Mean layer usage averages this fraction over the $D$ layers of each tokenizer. Perplexity is $\exp(-\sum_k p_k\log p_k)$, where $p_k$ is the assignment frequency of codeword $k$; suffix tokens are excluded.

\makeatletter\setlength{\@dblfpsep}{6pt}\makeatother

\begin{table}[!htbp]
\captionsetup{skip=4pt}
\caption{Full-catalog test results on Music4All-Onion (seed 2020). Random labels give input dimensions; ID-only SASRec is in Table~\ref{tab:id-only}. Bold/underlined values mark the best/second-best audio inputs per recommender and metric using unrounded scores. R/N/M denote Recall/NDCG/MRR at 50.}
\label{tab:full-catalog-results}
\centering
\footnotesize
\setlength{\tabcolsep}{0.6pt}
\renewcommand{\arraystretch}{1.03}
\begin{tabular*}{\columnwidth}{@{\extracolsep{\fill}}lrrrrrrrrr@{}}
\toprule
& \multicolumn{3}{c}{\textbf{KNN}}
& \multicolumn{3}{c}{\textbf{SASRec}}
& \multicolumn{3}{c}{\textbf{TIGER}} \\
\cmidrule(lr){2-4}\cmidrule(lr){5-7}\cmidrule(lr){8-10}
\textbf{Input}
& \multicolumn{1}{c}{\textbf{R}$\uparrow$}
& \multicolumn{1}{c}{\textbf{N}$\uparrow$}
& \multicolumn{1}{c}{\textbf{M}$\uparrow$}
& \multicolumn{1}{c}{\textbf{R}$\uparrow$}
& \multicolumn{1}{c}{\textbf{N}$\uparrow$}
& \multicolumn{1}{c}{\textbf{M}$\uparrow$}
& \multicolumn{1}{c}{\textbf{R}$\uparrow$}
& \multicolumn{1}{c}{\textbf{N}$\uparrow$}
& \multicolumn{1}{c}{\textbf{M}$\uparrow$} \\
\midrule
\multicolumn{10}{@{}l}{\textbf{Music4All-Onion-Full (56,512 tracks)}} \\
Wav2Vec2 & 0.0027 & 0.0007 & 0.0003 & 0.4467 & 0.2794 & 0.2360 & 0.2127 & 0.1470 & 0.1303 \\
Music2Vec & 0.0045 & 0.0013 & 0.0005 & 0.4471 & 0.2799 & 0.2365 & 0.2169 & 0.1508 & 0.1340 \\
HuBERT & 0.0031 & 0.0009 & 0.0004 & 0.4463 & 0.2793 & 0.2360 & 0.2134 & 0.1488 & 0.1325 \\
MERT & 0.0031 & 0.0008 & 0.0003 & \underline{0.4472} & \textbf{0.2804} & \textbf{0.2371} & 0.2314 & \underline{0.1575} & \underline{0.1388} \\
CLAP-G & \underline{0.0056} & \underline{0.0015} & \underline{0.0006} & \textbf{0.4475} & \underline{0.2801} & \underline{0.2366} & \underline{0.2364} & 0.1565 & 0.1364 \\
CLAP-Music & \textbf{0.0108} & \textbf{0.0030} & \textbf{0.0012} & 0.4469 & 0.2795 & 0.2362 & \textbf{0.2581} & \textbf{0.1696} & \textbf{0.1471} \\
\addlinespace[2pt]
Random-512 & $0.0008$ & $0.0002$ & $0.0001$ & $0.4461$ & $0.2803$ & $0.2372$ & $0.0203$ & $0.0059$ & $0.0025$ \\
Random-768 & $0.0007$ & $0.0002$ & $0.0000$ & $0.4457$ & $0.2806$ & $0.2378$ & $0.0969$ & $0.0574$ & $0.0474$ \\
\bottomrule
\end{tabular*}
\end{table}

\makeatletter\setlength{\@dblfpsep}{6pt}\makeatother

\makeatletter\setlength{\@dblfpsep}{6pt}\makeatother

\begin{table}[!htbp]
\caption{Reference-tokenizer statistics ($W=256,D=3$) for Table~\ref{tab:prefix}. Cells give active codewords / usage (\%) / perplexity.}
\label{tab:prefix-codebook-usage}
\centering\footnotesize
\setlength{\tabcolsep}{1pt}
\renewcommand{\arraystretch}{1.0}
\begin{tabular*}{\columnwidth}{@{\extracolsep{\fill}}lccc@{}}
\toprule
Embedding & Layer 1 & Layer 2 & Layer 3 \\
\midrule
\multicolumn{4}{@{}l}{\textbf{LFM-AM-4996}} \\
Wav2Vec2 & 17 / 6.64 / 7.97 & 80 / 31.25 / 27.42 & 95 / 37.11 / 10.25 \\
Music2Vec & 32 / 12.50 / 13.82 & 55 / 21.48 / 18.10 & 49 / 19.14 / 17.11 \\
HuBERT & 77 / 30.08 / 20.39 & 104 / 40.62 / 40.75 & 28 / 10.94 / 1.72 \\
MERT & 30 / 11.72 / 15.80 & 59 / 23.05 / 13.17 & 109 / 42.58 / 28.21 \\
CLAP-G & 75 / 29.30 / 35.20 & 144 / 56.25 / 49.48 & 126 / 49.22 / 28.04 \\
CLAP-Music & 28 / 10.94 / 16.88 & 28 / 10.94 / 1.95 & 34 / 13.28 / 6.68 \\
\midrule
\multicolumn{4}{@{}l}{\textbf{Onion-Dedup-5000}} \\
Wav2Vec2 & 82 / 32.03 / 38.00 & 94 / 36.72 / 35.20 & 122 / 47.66 / 35.03 \\
Music2Vec & 22 / 8.59 / 7.71 & 80 / 31.25 / 32.25 & 73 / 28.52 / 6.89 \\
HuBERT & 86 / 33.59 / 25.74 & 87 / 33.98 / 36.78 & 67 / 26.17 / 12.06 \\
MERT & 71 / 27.73 / 20.86 & 81 / 31.64 / 39.57 & 54 / 21.09 / 5.51 \\
CLAP-G & 34 / 13.28 / 10.43 & 54 / 21.09 / 10.91 & 33 / 12.89 / 4.99 \\
CLAP-Music & 50 / 19.53 / 15.68 & 71 / 27.73 / 13.99 & 52 / 20.31 / 3.10 \\
\bottomrule
\end{tabular*}
\end{table}

\begin{table}[!htbp]
\caption{ID-only SASRec test results at 50: mean $\pm$ sample SD over seeds 2020--2024 on the smaller catalogs; seed 2020 on Full. Dim. is item dimension, $n$ is seed count, and N denotes NDCG.}
\label{tab:id-only}
\centering\footnotesize
\setlength{\tabcolsep}{2pt}
\renewcommand{\arraystretch}{1.05}
\begin{tabular*}{\columnwidth}{@{\extracolsep{\fill}}rrrrr@{}}
\toprule
Dim. & $n$ & R@50 & N@50 & MRR@50 \\
\midrule
\multicolumn{5}{@{}l}{\textbf{LFM-AM-4996}} \\
512 & 5 & $0.4684\pm0.0024$ & $0.2895\pm0.0009$ & $0.2434\pm0.0009$ \\
768 & 5 & $0.4577\pm0.0024$ & $0.2854\pm0.0014$ & $0.2409\pm0.0020$ \\
\midrule
\multicolumn{5}{@{}l}{\textbf{Onion-Dedup-5000}} \\
512 & 5 & $0.4625\pm0.0031$ & $0.2776\pm0.0012$ & $0.2304\pm0.0015$ \\
768 & 5 & $0.4538\pm0.0007$ & $0.2729\pm0.0009$ & $0.2267\pm0.0011$ \\
\midrule
\multicolumn{5}{@{}l}{\textbf{Music4All-Onion-Full}} \\
512 & 1 & $0.4468$ & $0.2796$ & $0.2362$ \\
768 & 1 & $0.4473$ & $0.2814$ & $0.2383$ \\
\bottomrule
\end{tabular*}
\end{table}

\begin{table}[!htbp]
\caption{Recall@50 (R) and MRR@50 (M) for retained-layer validation experiments at epoch 10, complementing NDCG@50 in Table~\ref{tab:prefix}. Digits identify retained layers; bold and underlined values mark the best and second-best subsets using unrounded scores.}
\label{tab:prefix-full}
\centering
\footnotesize
\setlength{\tabcolsep}{0.6pt}
\renewcommand{\arraystretch}{0.90}
\begin{tabular*}{\columnwidth}{@{\extracolsep{\fill}}llrrrrrrr@{}}
\toprule
Embedding & Metric & P1 & P2 & P3 & P12 & P13 & P23 & P123 \\
\midrule
\multicolumn{9}{l}{\textbf{LFM-AM-4996}} \\
\multirow{2}{*}{Wav2Vec2} & R & \underline{0.4206} & 0.3844 & \textbf{0.4298} & 0.3846 & 0.3863 & 0.3688 & 0.3687 \\
 & M & \underline{0.2305} & 0.2201 & \textbf{0.2394} & 0.2167 & 0.2135 & 0.2107 & 0.2098 \\
\addlinespace[0.5pt]
\multirow{2}{*}{Music2Vec} & R & \textbf{0.4196} & 0.4113 & \underline{0.4127} & 0.3822 & 0.3616 & 0.3769 & 0.3726 \\
 & M & \textbf{0.2298} & 0.2277 & \underline{0.2287} & 0.2128 & 0.2075 & 0.2102 & 0.2055 \\
\addlinespace[0.5pt]
\multirow{2}{*}{HuBERT} & R & \textbf{0.4115} & 0.3784 & 0.3943 & 0.3712 & \underline{0.4073} & 0.3701 & 0.3735 \\
 & M & \underline{0.2279} & 0.2202 & \textbf{0.2476} & 0.2120 & 0.2251 & 0.2127 & 0.2116 \\
\addlinespace[0.5pt]
\multirow{2}{*}{MERT} & R & 0.3990 & \textbf{0.4271} & \underline{0.4020} & 0.3824 & 0.3724 & 0.4008 & 0.3636 \\
 & M & 0.2250 & \textbf{0.2348} & \underline{0.2300} & 0.2146 & 0.2102 & 0.2193 & 0.2038 \\
\addlinespace[0.5pt]
\multirow{2}{*}{CLAP-G} & R & \underline{0.3854} & 0.3756 & \textbf{0.4069} & 0.3496 & 0.3606 & 0.3548 & 0.3541 \\
 & M & \underline{0.2206} & 0.2151 & \textbf{0.2289} & 0.1998 & 0.2059 & 0.2029 & 0.2011 \\
\addlinespace[0.5pt]
\multirow{2}{*}{CLAP-Music} & R & 0.3948 & 0.4027 & \textbf{0.4330} & 0.4005 & 0.3832 & \underline{0.4222} & 0.3814 \\
 & M & 0.2205 & \textbf{0.2529} & 0.2308 & 0.2200 & 0.2129 & \underline{0.2321} & 0.2132 \\
\midrule
\multicolumn{9}{l}{\textbf{Onion-Dedup-5000}} \\
\multirow{2}{*}{Wav2Vec2} & R & 0.3729 & \textbf{0.3788} & \underline{0.3748} & 0.3589 & 0.3451 & 0.3466 & 0.3336 \\
 & M & 0.2071 & \textbf{0.2092} & \underline{0.2083} & 0.1969 & 0.1933 & 0.1947 & 0.1846 \\
\addlinespace[0.5pt]
\multirow{2}{*}{Music2Vec} & R & \underline{0.4194} & 0.3800 & \textbf{0.4279} & 0.3737 & 0.4055 & 0.3701 & 0.3750 \\
 & M & \underline{0.2207} & 0.2083 & \textbf{0.2284} & 0.1951 & 0.2121 & 0.2014 & 0.2018 \\
\addlinespace[0.5pt]
\multirow{2}{*}{HuBERT} & R & \underline{0.3753} & 0.3554 & \textbf{0.4187} & 0.3565 & 0.3513 & 0.3476 & 0.3396 \\
 & M & \underline{0.2066} & 0.1987 & \textbf{0.2257} & 0.1975 & 0.1946 & 0.1966 & 0.1878 \\
\addlinespace[0.5pt]
\multirow{2}{*}{MERT} & R & \underline{0.4038} & 0.3764 & \textbf{0.4217} & 0.3492 & 0.3764 & 0.3659 & 0.3417 \\
 & M & \underline{0.2191} & 0.2097 & \textbf{0.2269} & 0.1928 & 0.2055 & 0.2003 & 0.1861 \\
\addlinespace[0.5pt]
\multirow{2}{*}{CLAP-G} & R & \underline{0.4130} & 0.4038 & \textbf{0.4270} & 0.3888 & 0.4000 & 0.3914 & 0.3790 \\
 & M & \underline{0.2268} & 0.2142 & \textbf{0.2304} & 0.2075 & 0.2108 & 0.2017 & 0.1969 \\
\addlinespace[0.5pt]
\multirow{2}{*}{CLAP-Music} & R & \textbf{0.4208} & 0.4086 & \underline{0.4151} & 0.3836 & 0.3938 & 0.3979 & 0.3714 \\
 & M & \underline{0.2214} & 0.2134 & \textbf{0.2327} & 0.2039 & 0.2075 & 0.2113 & 0.1963 \\
\bottomrule
\end{tabular*}
\end{table}

\section{Control Protocols}
\label{app:control-details}
Tables~\ref{tab:main-results} and~\ref{tab:full-catalog-results} report random-feature results; Table~\ref{tab:id-only} reports ID-only SASRec. Each dimension-matched random bank is shared across recommenders and encoders. Audio and random TIGER use input-specific tokenizers, whereas assignment controls fix the CLAP-Music tokenizer and permute complete SID rows.

Controls retain the model settings in Appendix~\ref{app:configuration-details}, epoch 10, and last-100 exclusion. ID-only uses default item initialization without external vectors or row normalization. Random runs vary the feature-bank seed and train a fresh TIGER tokenizer per bank. Audio gains pair dataset, method, dimension, and seed; relative gains are $100(\overline{M}_{\mathrm{audio}}-\overline{M}_{\mathrm{control}})/\overline{M}_{\mathrm{control}}$ on matched seeds.

Assignment controls use $W=256,D=3,P=4096$, four slots, the fixed reference tokenizer, and the same test targets as the main comparison. Paired estimates match dataset and generator seed independently of the seven-subset validation study. Each run contributes once to its condition mean, and each complete four-condition seed block once to the interaction.

\section{Configuration Details}
\label{app:configuration-details}

\paragraph{Model and optimization settings}
Audio inputs follow Table~\ref{tab:audio-encoders}. KNN uses row-normalized audio vectors and the normalized mean of the full observed history, with no training. It computes exact float32 cosine scores over the catalog and retrieves the top 50 tracks.

SASRec uses two layers and four attention heads, a hidden size matching the audio dimension (768 or 512), a feed-forward size of 2048, GELU activations, hidden and attention dropout of 0.2, and a tied input/output item table. It is trained with Adam at learning rate $10^{-4}$ and zero weight decay, batch size 512, and full-catalog cross-entropy for 10 epochs. Training and scoring use FP32.

TIGER uses four encoder and four decoder layers, hidden size 128, eight attention heads, feed-forward size 512, and dropout 0.1. Training uses AdamW with learning rate $3\times10^{-4}$, weight decay 0.01, a constant schedule, batch size 256, and 10 epochs. The forward pass uses FP16 with gradient scaling, except for the LFM-AM-4996/HuBERT retained-layer runs, which use FP32; cross-entropy is computed in FP32. Constrained decoding uses FP32, beam size 200, and 20 expansions per beam.

The RQ-VAE encoder uses hidden sizes $(768,384,192)$ for 768-dimensional inputs and $(512,256,128)$ for 512-dimensional inputs, with latent size 32 and a mirrored decoder, without dropout or batch normalization. Training uses Adagrad~\cite{duchi2011adagrad} with learning rate 0.04 and batch size 512, MSE reconstruction and quantization loss with weight 1, and commitment coefficient 0.25. Codebooks use 50-iteration k-means initialization; Sinkhorn normalization uses $\epsilon=0.2$ and 50 iterations.

\paragraph{Configuration choice}
For TIGER, validation-stage development on Onion-Dedup-5000 and the full
catalog with Wav2Vec2 and HuBERT compared learning-rate/weight-decay pairs
$(3\times10^{-4},0.01)$, $(10^{-3},0)$, and $(3\times10^{-4},0)$.
The reported experiments use $(3\times10^{-4},0.01)$ across encoders.

\noindent SASRec uses a fixed learning rate of $10^{-4}$. Both recommenders use
the epoch-10 checkpoint, without early stopping or validation-best checkpoint
selection.

\paragraph{Tokenizer training and reuse}
Reference tokenizers ($W=256,D=3$) probe the largest identical-RQ-tuple bucket among the first 4,096 catalog tracks every 30 epochs, stopping when it falls below 256 or at 5,000 epochs. Sweep tokenizers train for 3,000 optimizer updates. Tokenizers remain fixed during generator training. Width replicates (seeds 2020--2024) reuse one tokenizer and SID set per condition; retained-layer experiments reuse one per dataset--encoder pair (84 conditions). Appendix~\ref{app:control-details} specifies the controls.

\clearpage
\end{document}